\documentclass[acmtog,  balance=false,  table, nonacm]{acmart}
\usepackage{xcolor}
\usepackage{cleveref}

\usepackage{wrapfig}
\usepackage{makecell}
\usepackage{titlesec}
\titlespacing*{\section}
{0pt}{1.5ex plus 1ex minus .2ex}{1.0ex plus .2ex}
\titlespacing*{\subsection}
{0pt}{1.5ex plus 1ex minus .2ex}{1.0ex plus .2ex}

\DeclareMathOperator*{\argmin}{argmin}

\DeclareMathOperator*{\tr}{tr}
\usepackage{mfirstuc}
\newcommand{\reffig}[1]{Fig.~\ref{fig:#1}}
\newcommand{\refeq}[1]{Eq.~(\ref{eq:#1})}
\newcommand{\refsec}[1]{Sec.~\ref{sec:#1}}
\newcommand{\reftab}[1]{Table.~\ref{tab:#1}}
\newcommand{\refapp}[1]{App.~\ref{app:#1}}
\newcommand{\mat}[1]{\mathbf{#1}}
\newcommand{\E}{\mathbb{E}}
\newcommand{\Pb}{\mathbb{P}}

\newcommand{\y}{\boldsymbol{y}}

\newcommand{\qoi}{U}

\newcommand{\D}{\mat{D}}

\providecommand{\d}{}
\renewcommand{\d}{\boldsymbol{d}}

\newcommand{\edit}[1]{{ #1}}

\newcommand{\f}{\boldsymbol{f}}

\newcommand{\Rn}[1]{\mathbb{R}^{#1}}

\newcommand{\F}{\mat{F}}
\newcommand{\R}{\mat{R}}
\providecommand{\S}{}
\renewcommand{\S}{\mat{S}}

\newcommand{\K}{\mat{K}}
\newcommand{\x}{\boldsymbol{x}}

\providecommand{\V}{}
\renewcommand{\V}{\mat{V}}

\providecommand{\N}{}
\renewcommand{\N}{\mat{N}}
\providecommand{\U}{}
\renewcommand{\U}{\mat{U}}
\providecommand{\B}{}
\renewcommand{\B}{\mat{B}}

\providecommand{\D}{}
\renewcommand{\D}{\mat{D}}
\providecommand{\H}{}
\renewcommand{\H}{\mat{H}}
\providecommand{\L}{}
\renewcommand{\L}{\mat{L}}
\providecommand{\M}{}
\renewcommand{\M}{\mat{M}}

\providecommand{\I}{}
\renewcommand{\I}{\mat{I}}

\providecommand{\xmodes}{}
\renewcommand{\xmodes}{Force-Dual modes }
\providecommand{\Xmodes}{}
\renewcommand{\Xmodes}{Force-Dual Modes}

\AtBeginDocument{%
  }

\acmISBN{978-1-4503-XXXX-X/18/06}

\begin{document}

\title{ \Xmodes: Subspace Design from Stochastic Forces}
\author{Otman Benchekroun}
\email{otman.benchekroun@mail.utoronto.ca}
\affiliation{University of Toronto\country{Canada}}

\author{Eitan Grinspun}
\affiliation{University of Toronto\country{Canada}}

\author{Maurizio Chiaramonte}
\affiliation{Meta Reality Labs\country{USA}}

\author{Philip Allen Etter}
\affiliation{Meta Reality Labs\country{USA}}

\renewcommand{\shortauthors}{Benchekroun et al.}

\begin{abstract}
Designing subspaces for Reduced Order Modeling (ROM) is crucial for accelerating finite element simulations in graphics and engineering. Unfortunately, it's not always clear which subspace is optimal for arbitrary dynamic simulation. We propose to construct simulation subspaces from force distributions, allowing us to tailor such subspaces to common scene interactions involving constraint penalties, handles-based control, contact and musculoskeletal actuation. To achieve this we adopt a statistical perspective on Reduced Order Modelling, which allows us to push such user-designed force distributions through a linearized simulation to obtain a dual distribution on displacements. To construct our subspace, we then fit a low-rank Gaussian model to this displacement distribution, which we show generalizes Linear Modal Analysis subspaces for uncorrelated unit variance force distributions, as well as Green's Function subspaces for low rank force distributions. We show our framework allows for the construction of subspaces that are optimal both with respect to physical material properties, as well as arbitrary force distributions as observed in handle-based, contact, and musculoskeletal scene interactions.  
\end{abstract}


\begin{teaserfigure}
   \centering%
    \includegraphics[width=\linewidth]{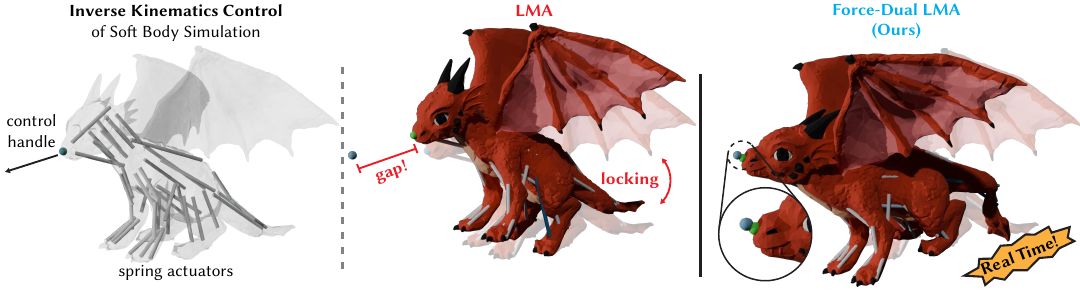}
    \caption{
We construct simulation subspaces by propagating user-defined or system-implied force distributions through the physical model, yielding a Force-Dual basis tailored to anticipated actuation patterns---such as the spring-driven inputs that lend themselves to walking, flapping, and wagging (\emph{left}). Compared to Linear Modal Analysis (LMA), which assumes spatially uncorrelated forces (\emph{middle}), our force-aware construction produces simulations that more accurately reflect localized, heterogeneous deformation, enabling real-time control of a 25,613-vertex, 103,307-tetrahedra deformable dragon running at 448 FPS (\emph{right}).
}
  \label{fig:teaser}
\end{teaserfigure}

\maketitle

\section{Introduction}
Whether it’s stretching before your early morning workout routine or tucking yourself into bed for a good night’s sleep, humans and animals alike are experts at interacting with the complex, deformable physical world. While we’ve become extremely good at  mimicking this rich physical detail in offline simulations for visual effects, our ability to interact with physically-based virtual environments in real time still remains primitive. This shortcoming stems from the fact that detailed soft body simulation entails solving an extremely high-dimensional optimization problem—where the solution, i.e., the displacement at each point in space, must be computed at every simulation timestep.

\begin{figure}[b]
    \centering
    \includegraphics[width=\linewidth]{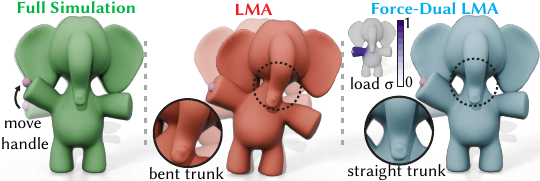}
    \caption{Interacting with this 8-dimensional Linear Modal Analysis subspace simulation of an elephant by moving its left arm upwards creates \textit{spooky action at a distance}, inducing deformation of the far away head and trunk. By revealing how to provide a better statistical prior to the subspace construction, we can avoid this artifact.}
    \label{fig:elephant_lodal_vs_modal}
\end{figure}

A popular approach to make this computation tractable for real-time interaction is Reduced Order Modeling (ROM), where a low-dimensional subspace for the displacements is precomputed, reducing the size of the optimization problem at runtime. However, it is notoriously difficult to construct subspaces that are well-suited to the broad and unpredictable range of interactions encountered during simulation. Each subspace typically captures only a narrow class of motions, and when a simulation strays outside this range, the resulting behavior can be physically implausible. One common artifact is \textit{spooky action at a distance}, where unrelated parts of the mesh appear to move in a correlated manner due to shared basis vectors (See \reffig{elephant_lodal_vs_modal}).

Part of the difficulty is that it is challenging to even characterize what kinds of motions a given subspace will be good at reproducing. Existing approaches often rely on manually selected or heuristically designed training motions, but there is no clear framework for systematically guiding subspace construction to match anticipated interactions.
We propose to address this by explicitly modeling the uncertainty inherent in interactive simulation through a statistical perspective, where subspace construction becomes equivalent to low-rank statistical model fitting of physical response data.

But how do we define the prior distribution from which this model fitting is performed? This is a central, unsolved question. While displacements are the primary quantities of interest in simulation, directly specifying distributions over them is unintuitive, as it requires prior knowledge of both material behavior and simulation dynamics.
Instead, we advocate for modeling distributions on the forces, which are typically easier to reason about, and then using a linearized dual relationship between forces and displacements to implicitly define a distribution over the latter. This shift enables a principled and flexible method for generating simulation subspaces.

\begin{figure}[t]
    \centering
    \includegraphics[width=\linewidth]{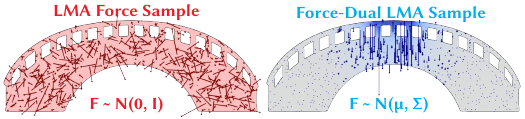}
    \caption{Traditional Linear Modal Analysis assumes the bridge to be subject to completely random uncorrelated Gaussian forces. We generalize Linear Modal Analysis by considering arbitrary force distributions, enabling us to construct subspaces better suited to a large variety of force interactions.}
    \label{fig:sample-force-bridge}
\end{figure}

We provide a mathematical framework for treating force distributions as first-class entities in subspace construction. This framework is general and expressive, encompassing prior approaches such as Linear Modal Analysis and Green’s functions, which can now be reinterpreted as arising from implicit statistical assumptions about the underlying force distributions, as shown in \reffig{sample-force-bridge}. These methods didn’t lack a statistical prior—they simply assumed one without making it explicit, often leading to brittle behavior when the actual forces diverge from those assumptions.

\edit{Our framework is the first to support the construction of subspaces that are not only tailored to the physical properties of the simulated material, but also customizable to general force distributions}. These distributions can be induced from various sources, such as example trajectories, actuation models, or interactive constraints, allowing users to easily design simulation behaviors that match their specific application needs. As a result, we can robustly simulate complex heterogeneous systems involving diverse force interactions, including contact, spring actuation, moving handle constraints and musculoskeletal actuation.

\section{Related Work}

The introduction of Linear Modal Analysis (LMA) to the computer graphics community by \citet{pentlandwilliams1989goodvibrations} brought with it a promise of fully interactive physics simulation through the use of linear subspaces.
Unfortunately, this promise didn't materialize because the resulting simulation's quality was noticed to degrade significantly as it underwent large deformation, specifically rotational motion. 

\subsection{Subspaces for Large Deformation}
Much work has been done on attempting to augment Linear Modal Analysis subspaces in order to allow  them to account for large rotational deformation. 

\begin{figure}
    \centering
    \includegraphics[width=\linewidth]{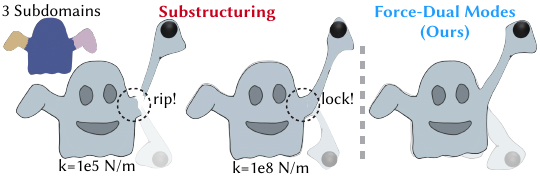}
    \caption{
    Localizing deformations via domain decomposition easily leads to ripping or locking artifacts based on the user's choice of domain coupling stiffness $k$. Instead, we interpret the same input subdomains as independent regions upon which Gaussian-distributed forces are applied.
    This correctly allows for deformation to leak between subdomains, leading to a simulation free from these artifacts. }
\label{fig:ghost-substructuring-comparison}
\end{figure}

Modal derivatives \cite{barbic2005realtimestvk} view the linearity of the subspace itself as the issue, and augment the linear subspace with \textit{derivative modes}, effectively making the subspace quadratic.
Alternatively, one can decompose the domain into various substructures, explicitly tracking the rotational motion for each component while also allowing every cluster to have its own separate linear subspace \cite{kim2011multidomain, barbic2005realtimestvk}. 
These unfortunately suffer from visual artifacts at substructure interfaces, as shown in \reffig{ghost-substructuring-comparison}.
\citet{choi2005modalwarping} instead explicitly account for rotational motion by rotating individual per-vertex components of the subspace itself, effectively doing domain decomposition at the finest per-vertex level.
Rotation strain coordinates \cite{huang2011RScoordantes, pan2015subspaceRS} decouple the rotation from the strain at a per-tet level, use a proper exponential integrator to advance the rotational component.

\citet{vontycowicz2013efficientreduceddeformableobjects} derive a column space expansion of a standard LMA subspace that significantly improves the amount of rotational motion it can capture.
Skinning eigenmodes \cite{benchekroun2023fastcomplementarydynamics, trusty2023subspacemfem} derive a linear subspace based off of linear blend skinning, enabling the capture of large rotational motion. %
These have since been extended with neural fields to be used for arbitrary geometric discretizations  \cite{changyue2023licrom, modi2024simplicits}.

All these methods only account for better accommodation of large rotational motion, however these methods do not account for the plethora of possible interactions the simulation may encounter at run-time.

\subsection{Subspaces for Interaction}

Aside from arbitrary large rotational deformation, a simulated soft body may undergo a large variety of other deformations at run-time.
What makes these other deformations particularly difficult to account for upon subspace construction is that we cannot a~priori be certain what the exact external forces the body will be exposed to throughout the interactive simulation. 
Nevertheless, different subspaces exist that make different assumptions on singular specific modes of expected user interaction. 

The first common approach is to 
One common approach is to construct subspaces for simulation based off of geometric posing and modeling techniques. 
\citet{pushkar2007harmoniccoordinatesforcharacterarticulation} create a subspace for posing their character from a user-designed coarse cage.
Instead of cages, one can also pose their character with skeleton rigs and design subspaces well tailored for smooth rig-based interaction \cite{jacobson2011boundedbiharmonicweights}.
Coupling these rig-posing spaces to a physics simulation is also possible \cite{hahn2012rigspacephysics, jacobson2012fastautomaticskinningtransformations, yu2015linearsubspacedesign}.

While these rig-space physics spaces excel at capturing bulk skeletal motion, they frequently lack fine scaled dynamic effects. 
To address this \citet{james2002DyRt} augment the controlled skeleton rig with modes obtained from LMA, and account for momentum transfer between the two. 
Because the LMA subspace falls out of date with large deformations, \citet{xu2016posespacesubspacedynamics} extend this and specifically build multiple LMA subspaces for various poses, and interpolate between them.
Instead of standard characters driven by an underlying articulated skeleton, another line of work adds fine-scale physical detail to arbitrary rig animations, by performing simulation in a rig-complementary subspace \cite{benchekroun2023fastcomplementarydynamics, zhang2020complementarydynamics, wu2023twowaycouplingskinningtransformationspbd}.

Aside from character posing, another mode of interaction that has been explored for subspace simulation is contact interaction.
A characteristic feature of contact simulation is its local response, which is the bane of any LMA-type subspace. 
To capture the local behavior, \citet{faure2011Sparsemeshlessmethods, brandt2018hyperreducedprojectivedynamics, lan2021medialipc} create a linear subspace with many sparsely sampled handles \cite{jacobson2011boundedbiharmonicweights}, allowing them to resolve the local response characteristic of a contact simulation.
\edit{
Accounting for uncertainty in the contact loads to which a fabricated object may be subjected, \citet{langlois2016stocasticfem} collect data from a rigid body simulation to construct a force distribution that is then mapped to a distribution on internal stresses. They use this stress field to predict probabilities of a fracture event occurring across the entire object.}

To represent arbitrary contacts that are encountered at run-time, many works turn to adaptive schemes that automatically introduce new locally-defined basis functions upon the detection of a new contact event. 
\citet{harmon2013subspaceintegration} augment the subspace with \textit{Boussinesq} functions about the contact point, while others bring back full order degrees of freedom at each vertex when enough local strain is detected as in \citet{yun2015subspacecondensation, trusty2024tradingspaces, mercier-aubin2022adaptiverigidification}.

Unfortunately, each of these subspaces is tailored to a specific mode of interaction comprising only a small range of the infinite possible interactions during an interactive simulation. 
For example, muscle actuation is a crucial mode of interaction with a physical environment, with many applications in character animation and biomechanics \cite{ lee2018dexterousmanipulationvolumetricmuscles}, however to the best of our knowledge, there are no subspace models that are specifically designed for muscle actuation. 
This leaves practitioners of these applications to make their own simplifications, bounding them to extremely coarse geometry in order to make their simulation tractable \cite{min2019softcon}.

\edit{We propose the first method for constructing simulation subspaces that are aware of arbitrary Gaussian force distributions, including those that arise from contact, handle interaction, as well as muscle, spring and pneumatic actuation.}

\section{Statistical Formalism for Model Reduction}

A core pillar of \xmodes is the reformulation of the model reduction problem into a probabilistic framework. While this perspective has deep roots in engineering—most notably through the Kosambi-Karhunen–Loève(KKL) expansion \cite{ghanem2003stochastic}—it is seldom explored in computer graphics, despite its potential to offer powerful insights for tasks we routinely face, such as evaluating the quality of reduced subspaces, quantifying approximation error under varying conditions, and reasoning about generalization to new inputs or unseen forces. 

At a high level, we may frame model reduction as approximating the outcome of a complex system governed by a computationally expensive procedure,
\begin{equation} \label{eq:modelreductionproblem}
\qoi(\omega) = \mathcal{L}(F(\omega)) ,
\end{equation}
where $\qoi(\omega) \in \mathbb{R}^n$ denotes the quantity of interest (e.g., displacements), and $F(\omega)\in \mathbb{R}^k$ represents the system’s input (e.g., boundary conditions, external forces, prior states).
We treat both $\qoi$ and $F$ as random variables dependent on a sample outcome $\omega \in \Omega$, capturing the natural variability and uncertainty present in future simulation configurations. We will suppress $\omega$ to simplify notation.

The assumption underlying model reduction is that the output $\qoi$ has a certain low-dimensional structure. 
For example, in linear model reduction, we approximately express the random structure of $\qoi$ as a small sum of linear components with amplitudes $Z \in \Rn{m}$ and an optional mean term $\mu_\qoi \in \Rn{n}$,
\begin{equation} \label{eq:pca}
    \qoi \approx  \B Z + \mu_\qoi \,,
\end{equation}
 where $\B \in \mathbb{R}^{n \times m}$ is the matrix whose $m$-th column corresponds to factor $Z_m$, and denotes the mean-zero error in the approximation. Given the basis $\B$, the optimal reconstruction of the factors $Z$ in the $\M$-norm is given by the projection
\begin{equation}
    Z = \B^\dag (\qoi - \mu_\qoi) \,,
\end{equation}
where $\B^\dag$ denotes the pseudo-inverse,
\begin{equation}
    \B^\dag = (\B^T \M \B)^{-1} \B^T \M^T \,.
\end{equation}

The objective of linear model reduction is to appropriately choose $\B$ such that the approximation error is minimized in an appropriate metric $\mat{M}$,
\begin{equation} \label{eq:minresidual}
    \min_{\B} \E  \| \qoi - \B \B^\dag \qoi \|_\mat{M}^2 \,.
\end{equation}

This problem, and related formulations, are common across many disciplines in engineering, statistics, and the sciences. In statistics, this framework is known as Principal Component Analysis (PCA) \cite{hotelling1933analysis}; in engineering, when reconstructing simulation snapshots, it is referred to as the Proper Orthogonal Decomposition (POD) \cite{sirovich1987turbulence}. Its continuous analog appears in stochastic mechanics as the KKL expansion \cite{ghanem2003stochastic}.

A unique solution to the problem does not exist unless one also applies an additional constraint. If one applies the $\mat{M}$-orthogonality constraint
\begin{equation}
    \mat{B}^T \mat{M} \mat{B} = \mat{I} \,,
\end{equation}
then it can be shown (see \refapp{deriving-optimal-subspace}) that the optimal choice of $\mat{B}$ are the \emph{top} $m$ eigenvectors of $\Sigma_\qoi \mat{M}$, i.e.,
\begin{equation} \label{eq:eigenvalue1}
    \Sigma_\qoi \mat{M} \mat{B} = \mat{B} \Lambda \,,
\end{equation}
where $\Lambda = \text{diag}(\lambda_1, .., \lambda_m)$ are the eigenvalues. And if $\Sigma_\qoi$ is invertible, then this is equivalent to the \emph{bottom} $m$ generalized eigenvectors of equation
\begin{equation} \label{eq:eigenvalue2}
     \Sigma_\qoi^{-1} \mat{B} = \mat{M} \mat{B} \Lambda^{-1} \,.
\end{equation}

To summarize, once we know a covariance $\Sigma_U$ for the quantity of interest, we can generate an error-minimizing subspace for simulation by solving the generalized eigenvalue problem \refeq{eigenvalue2}.
Different approaches—whether they acknowledge this explicitly or not—make different assumptions about this distribution.

Data-driven methods, such as POD, estimate $\Sigma_U$ from high-resolution simulation snapshots. While accurate, this process is often expensive, tedious, and problematic in interactive contexts. For example, it is paradoxical to predict the distribution of \textit{interactive} forces that will occur during interactive use when only relying on \textit{offline} simulations.
\begin{figure*}[t]
    \centering
    \includegraphics[width=\linewidth]{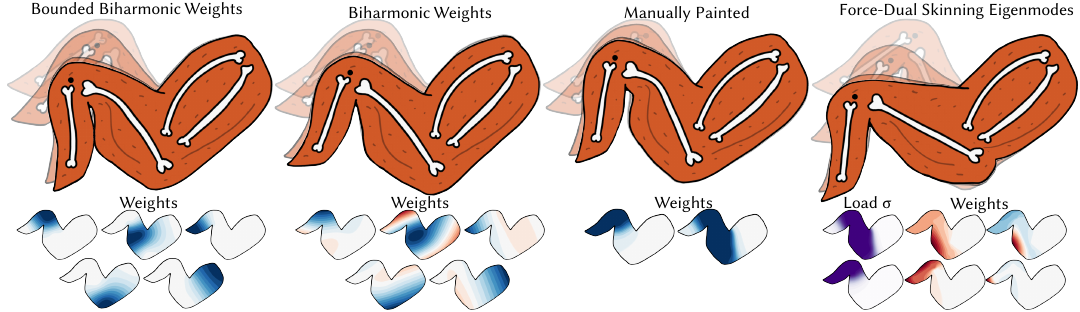}
    \caption{Interpreting user input as a force prior leads to a subspace that is physically aware, allowing us to fully bend the chicken wing backwards, without bending the bone. Instead, attempting to construct a subspace by directly prescribing a distribution on displacements leads to subspaces that are not material aware, and lead to locking during simulation, and a disrespect of the physical material properties.}
\label{fig:chicken_wing_force_distribution_vs_other_distributions}
\end{figure*}

Other approaches circumvent this by having users build smooth, localized weights \cite{yu2015linearsubspacedesign, jacobson2011boundedbiharmonicweights}. While this allows users to encode some intuitive features, such as the area of influence of interaction, it fails to capture critical aspects that are not easily conceptualized—such as boundary conditions, material heterogeneity, or anticipated contact regions (See \reffig{chicken_wing_force_distribution_vs_other_distributions}).

A third common choice is to derive $\Sigma_U$ from the inverse Hessian of an elastic energy function, which corresponds to Linear Modal Analysis (LMA) \cite{pentlandwilliams1989goodvibrations}. This approach is appealing due to its physical grounding, but it lacks any mechanism for incorporating prior knowledge about expected forces. For instance, the elastic energy Hessian does not encode information about where contact might occur, rendering it blind to one of the most important sources of localized deformation.

In this work, we propose a different approach: we construct $\Sigma_U$ not directly, but as the result of propagating uncertainty in the input forces $F$ through the system operator $\mathcal{L}$. In other words, we first model the statistics of $F$—which are often more intuitive and accessible to the user—and then transform those statistics through the system, imbuing them with non-intuitive physical information, to produce the resulting $\Sigma_U$.

As we will show, many interactions in a scene naturally induce plausible statistical models for $F$. These in turn give rise to a physically meaningful and computationally effective choice of covariance for the solution space. While users may struggle to guess the appropriate statistics of $\Sigma_U$ directly, they often have a much better intuition for the kinds of forces or interactions their system will encounter.

\section{Subspaces from Forces Distributions}
We will now derive our main contribution, a derivation of the covariance matrix $\Sigma_U$ from a distribution of forces. 
We formally define the process $\mathcal{L}$ for our application of elastodynamic simulation, and start at the standard variational perspective that casts simulation as the solution to a minimization problem,
\begin{align} \label{eq:least_action}
    U^* = \argmin_U \mathcal{V}(U) + \mathcal{C}(U),
\end{align}
where $\mathcal{V}$ represents a physical energy (e.g., elastic energy, kinetic energy, constraints) and $\mathcal{C}$ denotes an external influence on the system (e.g. from contact, actuation, or other user interaction). 
We make a first order approximation of our external potential to expose the external forces $F = -\nabla \mathcal{C}(U)$ acting on our system,
\begin{align} \label{eq:least_action_force_dual}
    U^* = \argmin_U \mathcal{V}(U) - U^T F.
\end{align}
Our goal is to relate the randomness in the distribution of $F$, to the distribution on $\qoi^*$.

When $\qoi$ is close to the rest state of the system at which $\mathcal{V}$ is minimized ($\qoi = 0$), one may perform a second order Taylor expansion of $\mathcal{V}$ to arrive at the quadratic approximation, $\mathcal{V}(\qoi) \approx \frac{1}{2} \qoi^T \mat{H} \qoi \,$, simplifying our variational problem further,
\begin{align} 
        U^* = \argmin_\qoi \frac{1}{2} \qoi^T \mat{H} \qoi - \qoi^T F,.\label{eq:linearized_least_action_force_dual}
\end{align}
where $\mat{H}$ is the physical energy Hessian at $\qoi = 0$.  
The solution to the above quadratic energy is then given by,
\begin{equation} \label{eq:linearized_duality}
    \qoi^* = \H^{-1} F \,.
\end{equation}
Assuming $F$ is a multivariate gaussian random variable $F \sim \mathcal{N} (\mu_F, \Sigma_F)$, the affine transformation of such a random variable is well known, and induces a Gaussian distribution on $\qoi \sim \mathcal{N}(\mu_{\qoi}, \Sigma_{\qoi})$, in terms of the parameters of the distribution of $F$, 
\begin{align}
\mu_\qoi &= \mat{H}^{-1} \mu_F  \,, \\
\Sigma_\qoi &= \mat{H}^{-1} \Sigma_F \mat{H}^{-1}  \,, \label{eq:statics_var_u}
\end{align}
where we've arrived at a distribution on displacements in terms of a distribution on forces. 

\subsection{Linear Modal Analysis as a Special Case} \label{sec:linear_modal}

LMA is a common technique used in structural analysis and model reduction that assumes that the most likely directions in which one will experience deformations is those with the minimum contribution to elastic energy, and therefore uses the lowest eigenvectors of the matrix $\mat{H}$ as a reduced basis,
 \begin{align} \label{eq:modal_analysis}
     \H \B_{\text{modal}} = \M \B_{\text{modal}} \mat{\Gamma} .
 \end{align}
Where $\mat{H}$, in this case is the elastic energy Hessian (also known as the stiffness matrix), $\mat{\Gamma}$ is a diagonal matrix of modal frequencies and $\mat{M}$ is the mass matrix. 
Within our Force-Dual framework, we show in \refapp{deriving-optimal-subspace} that by choosing our force covariance $\Sigma_F = \M$ to be the mass matrix, which arises from continuous white noise applied over our domain, we recover the exact same subspace as the one prescribed by LMA.
To the best of our knowledge, we are the first to formally characterize the exact force distribution assumed by LMA.

This analysis offers an alternative perspective on LMA that makes it clear why the resulting modes are almost always globally supported --- the forces $F$ are assumed to be uncorrelated and almost always nonzero everywhere. 
This global force prior assumption introduces artifacts that have plagued model reduction techniques for decades, such as \textit{spooky action at a distance}.
\subsection{Green's Functions as a Special Case}
\begin{figure}
    \centering
    \includegraphics[width=\linewidth]{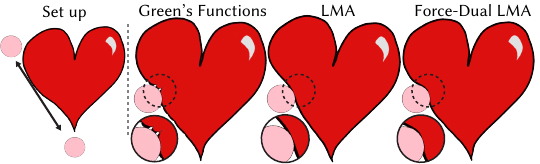}
    \caption{Green's functions describe responses from specific sampled sharp impulse forces, introducing artifacts when the observed forces are different from the samples.
    LMA constructs subspaces assuming spatially uncorrelated forces everywhere on the shape, struggling to accomodate specific local deformation.
    Our Force-Dual framework generalizes both of these approaches, and allows us to construct a subspace assuming a more realistic distribution of normal contact forces defined on the surface, resulting in a smooth localized response at run-time. }
    \label{fig:contact_heart}
\end{figure}

On the other extreme, instead of assuming random uncorrelated forces everywhere on the mesh, Green's functions form a subspace spanned by a small set of force-responses, a technique which has been used in structural analysis for thin plates and beams~\cite{kim2013nonlineardualmodes}.
While this perspective is useful when the \textit{exact} force impulses are known, the resulting subspace degrades in quality significantly when observed forces lie outside of its assumption, as shown in \reffig{contact_heart}, which uses a Green's function subspace constructed by sampling localized normal forces at random points on the left half of the heart.
We show in \refapp{greens_functions_from_force_dual_modes} that our Force-Dual modes framework also recovers Green's functions subspaces as a special case where the force distributions come from a set of randomly actuated forces, of smaller rank than the subspace dimension.

We view both LMA and Green’s function spaces as opposite extremes of force prior assumptions. At one end, LMA corresponds to maximal uncertainty: spatially uncorrelated forces of equal variance applied everywhere. At the other end, Green's function spaces correspond to minimal uncertainty: the set of possible force directions is known and fixed.

Our framework allows us to find a smooth middle ground between these two extremes, as shown in \ref{fig:contact_heart}, where our Force-Dual modes method constructs a prior of random uncorrelated normally directed forces on the left half of the heart. 
This lets us capture the local detail with more fidelity than LMA, without suffering from the sharp jarring deformations exhibited by the Green's function subspace, with the same number of degrees of freedom as both.

\section{Constructing Generalized Force Distributions}

Having established  that our method corresponds to LMA in the case where $F \sim \mathcal{N}(\mathbf{0}, \mathbf{M})$, we now take our framework beyond this narrow assumption and consider a more general distribution $F \sim \mathcal{N}(\mu_F, \Sigma_F)$.
As we will see, many modes of interaction are much better described by a different prior than the one assumed of LMA.
Properly making use of such priors is key in designing subspaces well suited to arbitrary forms of user interaction. 

Let us delve into two overarching strategies for constructing generalized force distributions.
We briefly begin with specifying nonuniform variances, before moving on to explore the construction of low-rank force distributions in greater detail.

\subsection{Variance Specification}

The simplest generalization for user interaction involves modifying only the diagonal components of $\Sigma_F$ with additional structure.
For example, by making the force distribution spatially localized, we can trivially generate subspaces that are localized as well. 
As an example, \reffig{batsy_local_force_dual_modes} shows that by localizing the force variance about the left wing of the bat, the resulting first three skinning eigenmodes reflect the locality of the deformation. 

On top of just locality, however,  our subspaces are also aware of properties of the PDE operator such as smoothness, material heterogeneity and imposed boundary conditions. %
This is because  our formulation makes use of the energy Hessian $\H$ in defining our final distribution,

If the domain is already split up into regions in which a user will interact, as is done in substructuring \cite{kim2011multidomain}, it's very easy to define subspaces based off of these regions by localizing a force distribution within that region as shown \reffig{ghost-substructuring-comparison}. 
Our subspace allows deformation to smoothly leak between adjacent substructures
without the need to tune auxiliary parameters such as interface stiffness.

Furthermore we find it is relatively easy to craft and \textit{paint} non-uniform force distributions that \textit{just work}, as we enable the user to focus on selecting the components of the shape they plan on handling independently, without having to worry about complex simulation details.
\reffig{chicken_wing_force_distribution_vs_other_distributions} shows a user crafting subspaces with various rig-weight crafting algorithms such as Bounded Biharmonic Weights \cite{jacobson2011boundedbiharmonicweights}, Biharmonic Coordinates \cite{yu2015linearsubspacedesign} and manual weight painting (guided by diffusion) with common geometry processing operations such as smoothing and diffusion.
All of these methods allow the bones to bend, failing to account for their prescribed near-rigid material stiffness.
By taking the exact same manually painted weights, and instead interpreting them as load variances, we can use our Force-Dual mode framework to obtain a Force-Dual subspace well suited to the heterogeneous flapping motion of the chicken wing.

\begin{figure}[t]
    \centering
    \includegraphics[width=\linewidth]{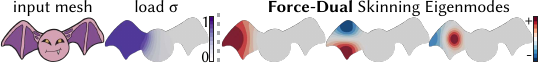}
    \caption{The locality of the input force distribution is reflected in the resulting force-dual skinning eigenmodes. }
\label{fig:batsy_local_force_dual_modes}
\end{figure}

\subsection{Linear Low-Rank Force Distributions}
Some simulation interactions imply their own specific force distribution, and capturing this force interaction is crucial for the quality of the resulting simulated interaction.
We assume these force interactions are linearly parameterized by a set of random actuation coefficients $A \in \Rn{r}$,
\begin{align}
    F = \D A,
\end{align}
where the columns of $\D \in \Rn{3n \times r}$ span the space of possible forces to which a simulation may be subjected.
It is then easier to reason reason about a Gaussian distribution on the actuation coefficients $A \sim \mathcal{N}(\mu_A, \Sigma_A)$, which induces a force distribution of 
\begin{align}
    F \sim \mathcal{N}(\D \mu_A, \D \Sigma_A \D^T).
    \label{eq:linear-low-rank-force-distribution}
\end{align}

In the following, we walk through several
strategies for constructing linear low-rank force distributions.

\begin{figure}[t]
    \centering
    \includegraphics[width=\linewidth]{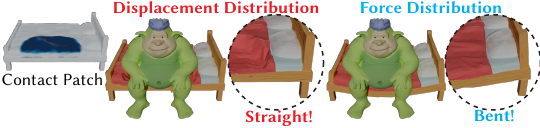}
    \caption{A contact patch being used to specify a distribution on displacements results in an extremely localized deformation, deforming only the mattress as the ogre sits. Interpretting the contact patch instead as specifying a force distribution allows the resulting simulation to also capture global components of the deformation, such as the bending of the stiff frame \refeq{contact_force_model}. In this example, contact simulation is performed using a standard quadratic spring penalty with a single mode, and the ogre is approximated with a spherical collider. }
    \label{fig:ogre-bed-contact}
\end{figure}

\paragraph{Point Handle Interaction}
A common interaction mode is to pin vertices and encourage them towards a target position via a quadratic spring-like objective.
\begin{align}
    \mathcal{C}(\qoi) = \alpha ||\S^T \qoi  - A ||^2_{\N}
\end{align}
with $\alpha$ being the strength of the objective,  $\S \in \Rn{3n \times r}$ is the selection matrix selecting vertex displacements from $\qoi$, $\N = \S^T \M \S$ is the diagonal mass-matrix containing the mass of the selected vertices, and $A \in \Rn{r}$ are the target displacements our selection should take. We again treat $A$ as normally distributed random variables.

The above objective leads to the forces 
\begin{align}
    F = - \frac{\partial \mathcal{C}}{\partial \qoi} = \alpha  \S \N A = \D A
\end{align}
whose form is such that we can use to \refeq{linear-low-rank-force-distribution} to define the final force distribution.
Note that if $A \sim \mathcal{N}(0, \textbf{I})$, the resulting $\Sigma_{\F}$ results in a diagonal distribution, where the only non-zero entries are those of the handle vertices, weighed by their mass and the penalty strength $\alpha$. Figs. \ref{fig:elephant_lodal_vs_modal}, \ref{fig:chicken_wing_force_distribution_vs_other_distributions}, \ref{fig:handle_adaptive_pegasus}, depict simulations involving point handle interaction with an elephant, chicken wing, and pegasus, respectively.

\paragraph{Contact}

A contact event $j$ applies a contact force at surface vertex $i$ as modelled by
\begin{align} 
    F_i &= \sum_j^{|C|}  v_i w_{ij} [\mat{n}_j \; \mat{t}_{j} \;\mat{r}_{j}]  
    \begin{bmatrix}
    A_{n_j} \\ 
    A_{t_j} \\
    A_{r_j}
    \end{bmatrix},  \label{eq:contact_force_model}
\end{align}
where $v_i$ is the mass of vertex $i$, $w_{ij}$ is a user crafted contact patch weight, defining the region on the mesh over which this force is presiding. The unit vectors $\mat{n}_j$, $\mat{t}_{j}$ and $\mat{r}_{j}$ are the contact normals, and two tangent vectors defining the frame of contact $j$.
The values $ A_{n_j},  A_{t_j},  A_{r_j}$ define the strength of each component of the contact force, which we treat as our normally distributed random actuations.
With this simple framework, we are able to model localized contact response with physically aware-properties as shown in \reffig{ogre-bed-contact}, where a user has crafted a contact patch for a single contact frame $w_j$.
Note that by using the Force-Dual modes framework, we get the expected buckling of the bed under the ogre's weight, whereas making use of just the user-crafted contact patch results in the overly localized mattress deforming, with the rest of the bed remaining immobile.

The above equation can be simplified into our canonical linear form, $F = \D A $, with $\D \in \Rn{3n \times 3|C|}$ where,
\begin{align}
    \D_{3i:3i+3, 3j:3j+3} &= w_{ij} [\mat{n}_j \; \mat{t}_{j} \;\mat{r}_{j}] \nonumber \; \mathrm{and} \\
    A_{3j:3j+3} &= [A_{n_j} \;  A_{t_j} \; A_{r_j}]^T. \nonumber
\end{align} 
From here, we can obtain the final distribution on forces from \refeq{linear-low-rank-force-distribution}.

\begin{figure}[t]
    \centering
    \includegraphics[width=\linewidth]{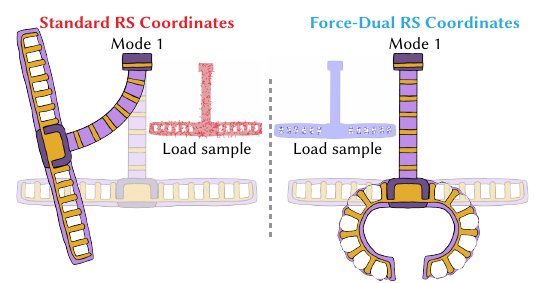}
    \caption{By leveraging our pneumatic actuation force prior, we correctly recover the expected behavior of the first mode, inflating the air pockets and bending the soft actuated gripper. In contrast, the first mode using the LMA force prior results in a completely uncharacteristic motion of the gripper. }
    \label{fig:pneumatic_actuation}
\end{figure}

\begin{figure}[t]
    \centering
    \includegraphics[width=\linewidth]{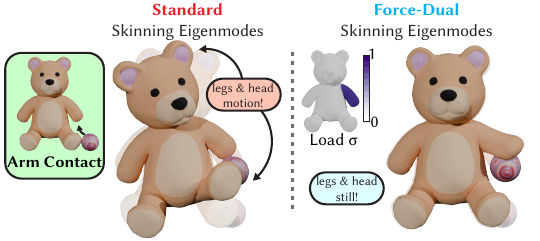}
    \caption{The ball collides against the teddy bear in a variety of different configurations. Our method allows us to construct load distributions for each contact configuration, and detect which of the distributions is active, allowing us to better capture the localized deformation induced by the contact at run-time. Using the standard force prior assumed with Skinning Eigenmodes leads to spooky action at a distance.}
    \label{fig:contact_bear}
\end{figure}

Although truly robust contact with large deformations usually requires an extremely non-linear incremental potential, making use of a linearized contact force approximation such as our own within the construction of our subspace greatly improves the run-time simulation quality, especially when compared to LMA and sparsely sampled Green's functions, as shown in \reffig{contact_bear} and \reffig{contact_heart}.
\paragraph{Pneumatic Actuation}
Pneumatic Elastic Actuation is an extremely popular actuation strategy in soft robotics \citep{xavier2022soft}, where air is pumped into an air pocket such that, upon inflation, causes an elastic force response corresponding to a desired bending motion.
Unfortunately, because the interaction of the air with the elastic material is crucial, modeling this interaction requires detailed expensive elastic simulation that cannot be carried out in real-time.
We show that by leveraging the force prior admitted by pneumatic actuation, we can build a subspace in which we can simulate such soft pneumatic actuation scenarios as shown in \reffig{pneumatic_actuation}.

Specifically, assuming $|P|$ air pockets, an air pocket $j \in \{ 0, 1, ... |P|-1\}$, applies a force to every surface vertex $i$ contouring that air pocket
\begin{align}
    F_i =  \begin{cases}
    v_i  \mat{n}_i A_j  &   i \in |V_j|  \\
    \boldsymbol{0} &  \mathrm{otherwise} \ ,
    \end{cases}
\end{align}
where $\mat{n}_i$ is the area averaged normal at vertex $i$, and $v_i$ is the Voronoi surface area of vertex $i$, and $|V_j|$ are all the surface vertices incident on pocket $j$. 
It is then easy` to recover our canonical low-rank force subspace, $F = \mat{D} A$, with $\mat{D} \in \Rn{3n \times |P|}$ where 
\begin{align}
    \mat{D}_{3i:3i+3, j} =  \begin{cases}
        v_i  \mat{n}_i   &   i \in |V_j| \\
        \boldsymbol{0} & \mathrm{otherwise } .
    \end{cases}
\end{align}

\begin{figure}[t]
    \centering
    \includegraphics[width=\linewidth]{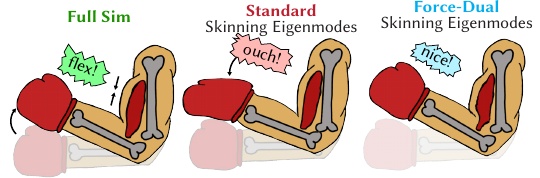}
    \caption{We can generate subspaces well suited for heterogeneous material biomechanical muscle actuation. Standard skinning eigenmodes are not aware of the intended user interaction, and cannot accommodate the simulated flexing of the bicep without visual artifacts. Both simulations use four skinning eigenmodes.
    \label{fig:arm-sim}}
\end{figure}

\paragraph{Muscle Actuation}
The capacity to simulate active musculoskeletons, such as those of animals and humans, remains the unfortunate computational bottleneck for many character control tasks with soft agents, as arise in soft robotics, medicinal geometry, or character animation.
This limits such applications to extremely coarse geometry \cite{min2019softcon, lee2018dexterousmanipulationvolumetricmuscles}.
We show in \reffig{arm-sim} that once again, by leveraging a simple force prior admitted from such actuated systems, we can enable interactive simulation of musculoskeletons that previously would have exhibited jarring visual physical artifacts. 
We adopt a simple and popular quadratic muscle actuation model \cite{ teran2003skeletalmuscle, modi2021emu} that defines an active muskuloskeletal energy.
Assuming we are given a set of $|T|$ active elements for our musculoskelature, we have 
\begin{align}
    \mathcal{C}(\qoi) = \frac{1}{2} \sum_j^{|T|} v_j \d^T_j \F^T_j \F_j \d_jA_j ,
\end{align}
where $A_j$ is the per-element actuation signal, $v_j$ is the per-element volume, $\d_j \in \Rn{3}$ is the per-element muscle fiber direction and $\F_j$ is the element deformation gradient.
The resulting force vector that arises, to first order, is given by
\begin{align}
    F &= - \frac{\partial \mathcal{C}}{\partial \qoi} = - \sum_t^{|T|} \frac{\partial \mathcal{C}}{\partial \F_j} : \frac{\partial \F_j}{\partial \qoi} , \\
    &= - \sum_j^{|T|}  v_j \F_j \d_j \d_j^T : \frac{\partial \F_j}{\partial \qoi}  A_j= \D A,
\end{align}
where $\D\in \Rn{3 |V| \times |T|}$ is a force actuation matrix mapping from active element actuation $A \in \Rn{|T|}$ to vertex forces and $ \frac{\partial \F_t}{\partial \qoi} \in \Rn{3 \times 3 \times 3 |V|}$ is the standard per-element Jacobian operator \cite{sifakis2012notes}.

\paragraph{Spring Actuation}
A similar type of actuation strategy popular in character animation and robotics are spring-based actuators \cite{geijtenbeek2013flexiblemuscle, tan2012softbodylocomotion, elatab2020soft}. 
These springs change their rest length during simulation in order to help the character or robot achieve a desired motion.

Starting from the external potential for a set of $r$ mass springs,
\begin{align}
    \mathcal{C}(U) = || \mat{l}(U) - \mat{l}_0 ||^2 ,
\end{align}
where $\mat{l} \in \Rn{r}$ and  $\mat{l}_0 \in \Rn{r}$ are both vectors of per-spring edge lengths. 
Taking the negative gradient with respect to $U$ reveals the induced force every timestep,
\begin{align}
    F =  \frac{\partial  \mat{l}}{\partial {U}}^T (\mat{l}_0 - \mat{l}(U)  ),
\end{align}
where we've revealed $(\mat{l}_0 - \mat{l}(U))$, the changing quantity that actuates our system. %
We choose this quantity to be modeled as a random set of actuation parameters, retrieving the canonical form required of \refeq{linear-low-rank-force-distribution}, 
\begin{align}
    F = \mat{D} A \ .
\end{align}
Where $\mat{D} =\frac{\partial  \mat{l}}{\partial {U}}^T  \in \R^{3n \times r}$ becomes the edge-length Jacobian.
\reffig{teaser} and \reffig{baby_dragon_varying_eigenmodes} show the real-time inverse kinematics control of a subspace simulation of a dragon.
Not leveraging this spring-actuator force prior leads to stiff and erroneously global motion that should be unachievable with the given spring actuator arrangement.

\section{Gaussian Mixtures for Multi-Modal Interaction}

Until now, we have assumed that a given force distribution prior is 
applicable throughout a simulation. However, simulations often have different regimes of actuation, for example an elastic system may be subject to a large localized contact force at one point in time, but later on may be dominated by more global inertial forces.
We can model such disparate modes of actuation by choosing a mixture of Gaussians for our force distribution, 
\begin{align}
    F = \sum_{k = 1}^{K} \chi_k F_k \ ,
\end{align}
where $F_k$ is the $i$-th Gaussian in our mixture model and the $\chi_k \in \{0, 1\}$ are mask variables such that the $k$-th Gaussian is only active when $\chi_k = 1$. 
The most common way to define a mixture is to let $\chi_k$ be the indicator for an independent categorical random variable $ C \in \{1, ..., K\}$,
\begin{equation}
    \chi_k = \begin{cases}
        1 & C = k \\
        0 & \text{otherwise}.
    \end{cases} 
\end{equation}
It is common to denote the probability distribution of $C$ as a vector $\pi$ such that $\pi_k = \mathbb{P}(C = k)$. The statistics of the mixture $F$ now read
\begin{equation}
\begin{split}
    \mu_F &\equiv \E[F] = \sum_{k = 1}^K \pi_k \mu_k \,, \\
    \Sigma_F &\equiv \text{cov}(F) = \E[\text{cov}(F \mid C)] + \text{cov}(\E[F \mid C]) \\
    &= \sum_{i = 1}^K \pi_k (\Sigma_k + (\mu_k - \mu_F)(\mu_k - \mu_F)^T) \,,
\end{split}
\end{equation}
where $\mu_k$ and $\Sigma_k$ are the means and covariances of the components $F_k$. 

\begin{figure}[t]
    \centering
    \includegraphics[width=\linewidth]{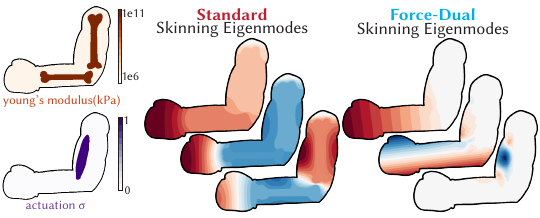}
    \caption{Our force-dual skinning eigenmodes reflect both the heterogeneous material properties of the material as well as the actuation distribution flexing the muscle fibers of the bicep. }
\label{fig:batsy_local_force_dual_modes}
\end{figure}

On top of allowing us to design subspaces for multiple modes of actuation, a mixture formulation allows us to have a probabilistically motivated way of adaptively selecting a subspace. 
In particular, if we knew which component $F_k$ was active during the simulation, we could select a basis of modes tailored for that subspace and solve for the amplitudes of those modes on-the-fly.

Although a direct observation of the force sample $F = \hat{F}$ does not uniquely identify which mixture component generated it, Bayes' Rule provides a principled way to infer the posterior probability of a component's responsibility for the observation during runtime,
\begin{equation}
    \Pb(C = k \mid F = \hat{F}) = \frac{\Pb(F \in dF(\hat{F}) \mid C = k) \, \Pb(C = k)}{\Pb(F \in dF(\hat{F}))}  \,. \footnote{Here, the notation $F \in dF(\hat{F})$ is necessary because technically the probability $F = \hat{F}$ is zero, so the above expression must be taken in the limit of $F$ being in an infinitesimal interval of size $dF$ around $\hat{F}$. }
\end{equation}
Taking logarithms and expanding,
\begin{equation} \label{eq:logbayes}
\begin{split}
    &\log \Pb(C = k \mid F = \hat{F}) \\
    &= -\frac{1}{2} (\hat{F} - \mu_k)^T \Sigma_k^{-1} (\hat{F} - \mu_k) - \frac{1}{2} \log \det(\Sigma_k) - \log c \,,
\end{split}
\end{equation}
\begin{figure}[b]
    \centering
    \includegraphics[width=\linewidth]{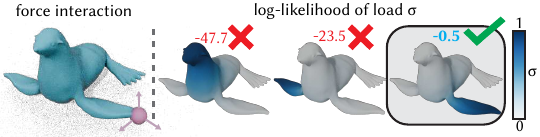}
    \caption{By leveraging Bayes' rule \refeq{logbayes}, we identify which force distribution was most likely to generate an observed sample force.}
    \label{fig:seal-maximum-likelihood}
\end{figure}
where $c$ is a constant. Therefore, from an observation $\hat{F}$ of forces acting on the object during runtime, we can adaptively select the most likely component that corresponds to this observation by finding the $k$ that maximizes \refeq{logbayes}. This allows us to choose our subspaces adaptively during the simulation by using a basis computed from the covariance matrix of the $k$-th component.
While truly evaluating \refeq{logbayes} is expensive as it requires inverting the $3n \times 3n$ covariance matrix of each k-th component of the force distributions, we often don't require computing the full inverse.
Because many of the observed interaction forces are usually only applied over a small sub-region of the domain, it's possible to marginalize the distributions on $F$ to focus on these sub-regions.
The resulting mean and covariance matrices of each Gaussian component can then be restricted to that given sub-region and quickly queried for real-time evaluation of \refeq{logbayes}.

\begin{figure}[t]
    \centering
    \includegraphics[width=\linewidth]{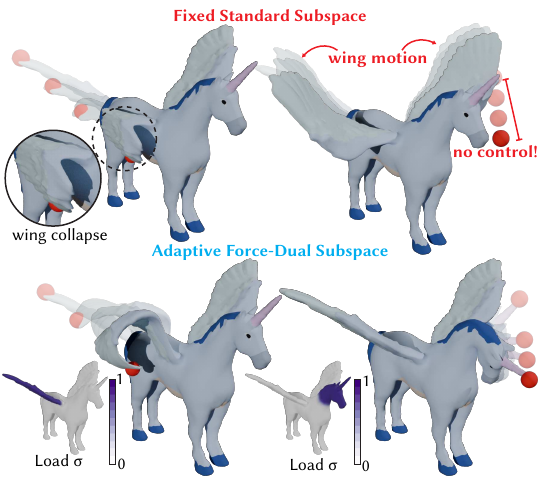}
    \caption{A user clicks on the pegasus to assign a control handle, and controls the pegasus by moving the control handle with a spring-like energy. Using only 2 skinning eigenmodes, but adaptively picking \textit{which two} modes based off the observed contact force, allows us to extremely efficiently model the control interactions on the pegasus.
    Using an equivalent two standard skinning eigenmodes without adaptivity results in the wing collapsing (rather than bending), as well as a loss of control when attempting to move the horn, which is instead met with a confusing and extreme global motion of the wings. }
    \label{fig:handle_adaptive_pegasus}
\end{figure}

\reffig{seal-maximum-likelihood} shows how given a selection of three load distributions defined on the seal, as well as an observed force, we can quantify how likely it was that the observed force was sampled from each of the load distributions, as marginalized over the subdomain on which the force is applied.
We make use of this ability to efficiently choose which subspace to use for simulation at run-time, for a scene with multimodal contact interaction \reffig{contact_bear}, and a scene with multimodal click interaction \reffig{handle_adaptive_pegasus}.
This allows us to capture, using only any two skinning eigenmodes at any point in time, a much more realistic and interactable simulation than using only a fixed subspace. 

\section{Implementation}
\label{sec:implementation}
The Generalized Eigenvalue Problem described by \refeq{eigenvalue1}, when solved naively, quickly becomes intractably expensive due to the dense computation required of
$\mat{H}^{-1}$.
To address this challenge, we adopt two different strategies for constructing our subspace depending on the structure of $\Sigma_F$.

The simplest case lies when $\Sigma_F$ is of full rank and either diagonal or block diagonal (with blocks small enough to invert quickly). 
In this scenario, we instead solve the equivalent problem of \refeq{eigenvalue2}, which finds the smallest $m$-eigenvectors of $\Sigma_U^{-1}$. 
Recalling our expression for the covariance on displacements $\Sigma_U^{-1} = \mat{H} \Sigma_F^{-1} \H$, the GEVP becomes
\begin{align}
    \mat{H} \Sigma_F^{-1} \H = \M \B \Lambda^{-1} \ .
\end{align}
Because the inverse $\Sigma_F^{-1}$ remains diagonally sparse and is fast to compute ($O(n)$) and $\mat{H}$ and $\mat{M}$ are both sparse as well, we can solve this problem with an out-of-the-box sparse GEVP solver, namely \texttt{eigs} from scipy \cite{scipy}.

Another common scenario arises when $\Sigma_F$ is not diagonal, and may contain frequent off-diagonal terms. 
In this scenario, one saving grace occurs if $\Sigma_F$ is low-rank.
In this event, solving for the subspace is still feasible if we have a Cholesky-like decomposition on $\Sigma_F$ and $\M$ of the form
\begin{align}
    \Sigma_F =  \mat{L} \mat{L}^T \quad \mathrm{and} \quad   \mat{M} = \mat{N}  \mat{N}^T  \ , \nonumber 
\end{align}
where $\L \in \Rn{3n \times r}$ and $\N \in \Rn{3n \times 3n}$ is sparse.
We return to the original form of \refeq{eigenvalue1} and introduce a change of variables on $\tilde \B = \mat{N}^T \B$.

With this in hand, we compute our subspace by performing a singular value decomposition on the matrix
\begin{align}
    \N \mat{H}^{-1} \L = \U \S \V^T \ ,
\end{align}
where we retain the top $m$ left-singular vectors and $m$ singular values in order to form our subspace matrix and eigenvalues
\begin{align}
    \B = \N^{-1} \U \quad \mathrm{and} \quad   \Lambda = \S^2 \ .
\end{align}
This strategy is particularly advantageous when $\L$ is thin ($r < 3n)$, as it only requires the solution to $r$ systems of equations sharing the same system matrix $\mat{H}^{-1} \L$, followed by a second set of linear solves $\B = \N^{-1} \U$.
Fortunately in our case, we always assume a lumped diagonal mass matrix, which makes both identifying $\N$ and inverting it trivial.
Additionally, in all our examples with specialized force distributions, we obtain the Cholesky-like factor for free, as it is given by the structure of the canonical force distribution itself and we can safely set $\L = \D$.

\section{Experiments and Discussion}

\subsection{Subspace Simulation Solvers}
Our method is extremely general, and can be made to accelerate many standard solvers popular in computer graphics.

By default, all examples shown in this paper employ a standard Newton solver with backtracking line search \cite{li2019optimizationtimeintegrator}, as this can be used to minimize any smooth energies with a first and second derivative, such as those that arise in contact \cite{Li2020IPC} or muscle actuation \cite{modi2021emu}.
To accelerate the evaluation of the reduced-space Hessian and gradient, we make use of a standard cubature scheme integration scheme \cite{an2008optimizingcubature}, where cubature points and weights are selected according to the method of \citet{trusty2023subspacemfem}.

For certain convex energies, it's possible to use a Projective Dynamics (PD) \cite{bouaziz2014projectivedynamics, brandt2018hyperreducedprojectivedynamics} solver to solve the optimization problem.
These solvers avoid recomputing the Hessian and drastically accelerate simulation, as shown in \reftab{statistics}.
For examples involving actuated spring systems, we use this solver with an ARAP \cite{kim2022dynamicdeformables} elastic energy for the volumetric deformation and the fast mass spring energy \cite{liu2013fastsimofmasssprings} for the spring actuator deformation. To accelerate the evaluation of the local step for the ARAP term, we make use of the clustering scheme of \citet{jacobson2012fastautomaticskinningtransformations}.

\subsection{Run-Time Analysis}
\reftab{statistics}  show a table with run-time analysis of some of the scenes presented in this paper, using both the Newton solver, or the PD solver described above, showing in brackets on the last column the speedup factor we achieve when compared to the full space. 
Our subspace simulations are consistently orders of magnitude faster than their full space counterparts, allowing for hard-real time interaction. 

\begin{table*}
\rowcolors{2}{white}{cyan!25}
\caption{ Run-time analysis for some large scenes presented in this paper.
\label{tab:statistics}
}
\begin{tabular}{|c|c|c|c|c|c|c|c|c|c|c|}
\hline
\textbf{Mesh} & \textbf{\makecell{\#Vertices}} & \textbf{\makecell{\# Tets}}  & \textbf{m } & \textbf{k}  & \textbf{Solver}  & \textbf{\makecell{Max \\ iters}} & \textbf{\makecell{Full Sim. \\ (sec/step)}} & \textbf{\makecell{FDM Sim. \\ (sec/step)}} & \textbf{\makecell{FDM Recon.\\ (sec/step)}} &  \textbf{\makecell{FDM Total \\ (sec/step)}} \\
\hline
Dragon (\reffig{teaser}) & 25,613 &  103,307 &  11 & 30  & PD & 10 &   $7.6 \times 10^{0}$  & $2.1 \times 10^{-3}$  & $1.3 \times 10^{-4}$ &  $2.4 \times 10^{-3}$ (\textbf{$\mathbf{\times 3166}$})\\
Teddy (\reffig{contact_bear}) &  21,209 &  89,942 & 24 & 50 &   Newton & 3  &  $6.7 \times  10^{0}$ &   $2.3 \times  10^{-2}$   & $2.6 \times  10^{-3}$ &  $2.7 \times 10^{-2}$ (\textbf{$\mathbf{\times 248}$}) \\
Pegasus (\reffig{handle_adaptive_pegasus})  &  15,281 & 58,401 & 24 & 200 &  Newton & 1 & $2.8 \times 10^0$ &  $3.2 \times 10^{-2}$ &   $1.9 \times 10^{-3}$&   $3.4 \times 10^{-2}$ (\textbf{$ \mathbf{ \times 82.4}$})\\
\hline
\end{tabular}
\end{table*}

\subsection{Comparison to Data-Driven Techniques}
Another option of estimating distributions on displacements is through data driven techniques such as the Proper Orthogonal Decomposition \cite{sirovich1987turbulence}, where offline simulations are recorded and the subspace is then constructed from performing Principal Component Analysis(PCA) on the interaction responses.

\begin{figure}
    \centering
    \includegraphics[width=\linewidth]{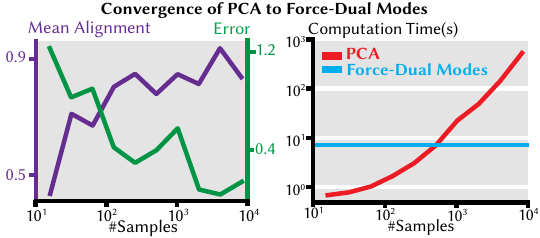}
    \caption{ A data-driven PCA subspace with enough force data samples theoretically converges to the same result as ours. We show that as more samples are collected, the first ten data-driven modes become better aligned to ours (top-left), as well as better reconstruct ours (top-right), with some variability due to sampling error. Increasing the observed samples for accuracy isn't free, and comes with a steeply increasing computational cost in the dense SVD at the heart of PCA (bottom). Our method does not require sampling and immediately recovers the true principal components of the desired distribution.}
    \label{fig:pca-vs-force-dual-modes}
\end{figure}

\reffig{pca-vs-force-dual-modes} shows that, in the limit of exhaustive samples, our two approaches agree on the same subspace at the same subspace, but ours arrives at it directly whereas a data-driven technique needs a lengthy training process to arrive at it. 
Unfortunately, these data-driven techniques are ultimately dependent on the data given to them, and collecting these samples can be an expensive, tedious process.

Our Force-Dual subspace construction framework shows that in the case where we can describe the force distributions with a Gaussian, we don't need to sample this process whatsoever, and can arrive directly at the closed form expression for the subspace.

\subsection{Ablation on Spatial Decay of Load Variance}
The more representative a force distribution to the true distribution of the observed forces, the more accurate the resulting simulation. 
\reffig{bear-convergence-comparison} explores the convergence of a subspace simulation to full space solution as the assumed force prior becomes more representative of the actual imposed load.

The ground truth solution is defined as the response to a localized load applied on the left arm of the bear.
We construct 4 different Force-Dual LMA subspaces, all assuming force priors that vary in their spatial decay and analyze how well they reconstruct the ground truth response as we vary the number of modes.
Specifically, the load variance of each component of the force distribution at a point $\x$ is given by $\sigma(\x) = 1 - (\frac{1}{\mathrm{exp}(-\alpha (||\x - \y|| - r)^2}) $, where we set $\alpha=10$ for a sharply decaying variance beyond radius $r$, which in turn controls the spatial locality of the variance.

For each radius value $c$, we compute the reconstruction error,
\begin{align}
    E_c = || \mat{u}^* - \mat{u}_c ||^2_\mat{M},
\end{align}
between the ground truth solution $\mat{u}^* = \mat{H}^{-1} \f \in \Rn{3n}$ and the reduced-space solution at that radius value $\mat{u}_c = \B_c (\B_c^T \H \B_c)^{-1} \B^T_c \f \in \Rn{3n}$.
\begingroup %
\setlength{\columnsep}{0.1 \linewidth}
\setlength{\intextsep}{0.05pt}
\begin{wrapfigure}{r}{0.4\linewidth}
\hspace*{-2.5em} \includegraphics[]{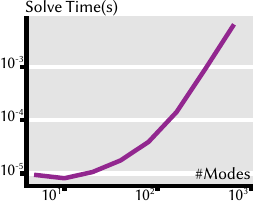}
\end{wrapfigure}
We show that even with a modestly localized force prior, we observe errors an order of magnitude smaller than the naive LMA baseline. 
Achieving lower error at low mode counts is especially important because the solution to a reduced space linear system evolves cubically with the size of the subspace, and simply adding more modes is rarely a sustainable strategy. 
The inset shows the solution time in seconds for a single reduced linear solve, of which frequently many need to be performed every non-linear simulation iteration.

\begin{figure}
    \centering
    \includegraphics[width=\linewidth]{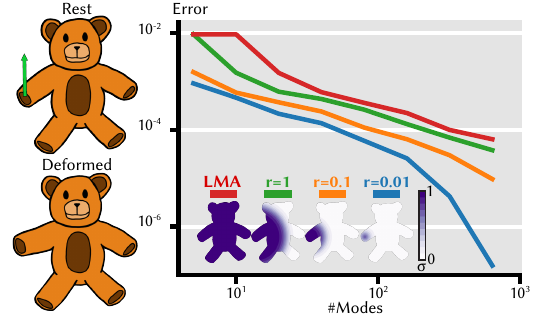}
    \caption{Our Force-Dual modes framework allows us to easily construct subspaces better suited for exotic force distributions. As the prior force distribution assumed more closely matches the one observed, the resulting error of our error results is orders of magnitude smaller than the standard one assumed of Linear Modal Analysis.}
    \label{fig:bear-convergence-comparison}
\end{figure}
\endgroup

\subsection{Non-linear Motion}
Our method, like Linear Modal Analysis, is developed on the linearized version of the variational form of the physical system. %
Despite this, our method can leverage decades of work in engineering and animation that enhances linear model reduction to better accommodate large non-linear deformation. 

\reffig{pneumatic_actuation} shows we can easily apply the Rotation Strain Coordinates \cite{huang2011RScoordantes} interpolation framework to properly reconstruct the large deformation associated with the soft pneumatic gripper.
Our framework requires no additional changes to be made to the Rotation Strain Coordinates algorithm, as they simply take as input a linear actuation and properly integrate the rotational component the actuation attempts to describe.

Furthermore, for many experiments involving large deformation, such as \reffig{ghost-substructuring-comparison}, \reffig{chicken_wing_force_distribution_vs_other_distributions}, \reffig{handle_adaptive_pegasus} and \reffig{bear-convergence-comparison}, we extend our framework and develop a Force-Dual version of Skinning Eigenmodes \cite{benchekroun2023fastcomplementarydynamics}.

To achieve this, we collapse the vector distributions on $U \in \Rn{3n}$ into a scalarized distribution on $\phi \in \Rn{n}$ by simply summing the individual components on $U$ 
\begin{align}
    \phi = U_x + U_y + U_z  \ .
\end{align}
The statistics on $\phi$ are now described by
\begin{align}
    \Sigma_\phi &= \Sigma_{U_{xx}} + \Sigma_{U_{yy}} + \Sigma_{U_{zz}} \nonumber, \\
    \mu_\phi &= \mu_{U_{x}} + \mu_{U_{y}} + \mu_{U_{z}} \ . \nonumber
\end{align}
We can then identify the principal components of this distribution, and use those as skinning weights to construct the same linear blend skinning subspaces as defined in \cite{benchekroun2023fastcomplementarydynamics}.
For low-rank force distributions, we still proceed as before in section \refsec{implementation}, assembling the factor $\K = \M \H^{-1} \L$.
However we instead perform an SVD on its scalarized $\K_\phi$, 
\begin{align}
    \K_\phi = \K_x + \K_y + \K_z \ ,
\end{align}
where $\K_d \in \Rn{n \times r}$ is the component of $\K$ corresponding to dimension $d$ only.
This extension recovers the same GEVP as developed by \citet{benchekroun2023fastcomplementarydynamics} when the assumed forces are Gaussian and uncorrelated $\Sigma_F = \M$.

\begin{figure}[t]
    \centering
    \includegraphics[width=\linewidth]{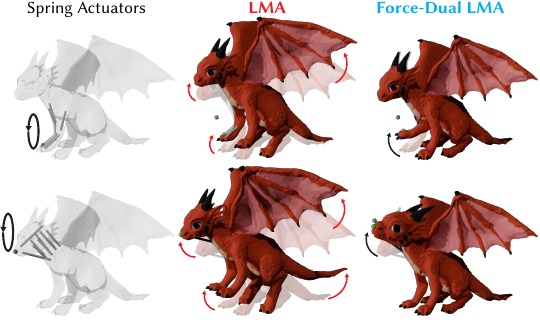}
    \caption{A single geometry may have arbitrary force distributions acting upon it. Ignoring this valuable prior and assuming the generic LMA force distribution leads to an extremely stiff simulation, where the dragon is unable to properly move its front leg (top), or head (bottom), without erroneously moving its entire body. Properly taking these distributions into account is crucial for realistic control tasks. }
\label{fig:baby_dragon_varying_eigenmodes}
\end{figure}

\section{Inverse Kinematic on Mass Spring Systems}
We are the first to show a system capable of real-time inverse kinematics on soft deformable characters of arbitrary mesh resolution with spring-like actuators, as shown in \reffig{teaser} and \reffig{baby_dragon_varying_eigenmodes}.
To achieve this we use a Levenberg-Marquardt solver \cite{nocedal1999numerical} to find the spring actuations, taking one step of this solver every simulation timestep.

We find that being explicit about properties of your load distribution upon subspace construction is \textit{critical} in physical control problems.
Because naive subspaces like LMA couple motion of far away vertices together, using them to solve control tasks allows them to \textit{cheat}.

For example, in \reffig{baby_dragon_varying_eigenmodes}, the inverse kinematics attempts to find the spring actuations of the head, or the front leg, that move a control point to a target location.
Using an LMA subspace lets the dragon attempt to satisfy this constraint by unnaturally moving its back legs and body forward, two locations where it does not have spring actuations and therefore should have no control over.
Our Force-Dual modes perspective emphasizes the importance of explicitly stating the underlying force distribution assumed of the subspace, resulting in a controlled simulation that naturally actuate the parts of the mesh the musculature can control.

\section{Limitations \& Conclusion}
We are extremely excited about many future directions towards which this research might lead.
In particular, the theory we have developed assumes linearized dynamics, making it easy to transfer between force distributions and displacement distributions. 
For simulations that exhibit extreme non-linearity, as is very common in highly articulated systems undergoing contact \cite{sharp2023datafreelearningreducedorderkinematics}, we hope to develop a non-linear version of this same theory, which would allow us to craft non-linear subspaces sensitive to non-linear actuation distributions.

Although we have primarily applied our framework to elastodynamics, the variational theory underpinning it is very general, and could be used to accurately accelerate  many other simulations subjected to well known force distributions, such as fluid control \cite{cui2018laplacianeigenfluids, chen2024fluid}.

Finally, the biggest bottleneck of soft-body character control remains the computational cost of the simulation. 
We think our Force-Dual subspace construction framework will be essential for making these previously impossible control problems feasible, and we look forward to attempting to use it on more complex control tasks such as swimming \cite{min2019softcon} or walking \cite{tan2012softbodylocomotion}.

In conclusion, we have shown that by leveraging a priori the expected distribution of forces a simulation will undergo, we can greatly improve the physical fidelity of interactive subspace simulations. 
To achieve this, we have presented a derivation of the subspace construction problem from a statistical perspective, allowing us to reason about fitting subspaces to specific force priors. 
Not only does this derivation generalize prior subspace construction techniques, it allows us to design novel subspaces well suited for various common scene interactions previously impossible, such as handle-based control, contact, musculoskeletal actuation and pneumatic actuation and spring actuation.

\bibliographystyle{ACM-Reference-Format}
\bibliography{sample-base}


\begin{thebibliography}{52}


\ifx \showCODEN    \undefined \def \showCODEN     #1{\unskip}     \fi
\ifx \showDOI      \undefined \def \showDOI       #1{#1}\fi
\ifx \showISBNx    \undefined \def \showISBNx     #1{\unskip}     \fi
\ifx \showISBNxiii \undefined \def \showISBNxiii  #1{\unskip}     \fi
\ifx \showISSN     \undefined \def \showISSN      #1{\unskip}     \fi
\ifx \showLCCN     \undefined \def \showLCCN      #1{\unskip}     \fi
\ifx \shownote     \undefined \def \shownote      #1{#1}          \fi
\ifx \showarticletitle \undefined \def \showarticletitle #1{#1}   \fi
\ifx \showURL      \undefined \def \showURL       {\relax}        \fi
\providecommand\bibfield[2]{#2}
\providecommand\bibinfo[2]{#2}
\providecommand\natexlab[1]{#1}
\providecommand\showeprint[2][]{arXiv:#2}

\bibitem[An et~al\mbox{.}(2008)]%
        {an2008optimizingcubature}
\bibfield{author}{\bibinfo{person}{Steven~S. An}, \bibinfo{person}{Theodore Kim}, {and} \bibinfo{person}{Doug~L. James}.} \bibinfo{year}{2008}\natexlab{}.
\newblock \showarticletitle{Optimizing cubature for efficient integration of subspace deformations}.
\newblock \bibinfo{journal}{\emph{ACM Trans. Graph.}} \bibinfo{volume}{27}, \bibinfo{number}{5}, Article \bibinfo{articleno}{165} (\bibinfo{date}{Dec.} \bibinfo{year}{2008}), \bibinfo{numpages}{10}~pages.
\newblock
\showISSN{0730-0301}
\urldef\tempurl%
\url{https://doi.org/10.1145/1409060.1409118}
\showDOI{\tempurl}


\bibitem[Barbi\v{c} and James(2005)]%
        {barbic2005realtimestvk}
\bibfield{author}{\bibinfo{person}{Jernej Barbi\v{c}} {and} \bibinfo{person}{Doug~L. James}.} \bibinfo{year}{2005}\natexlab{}.
\newblock \showarticletitle{Real-Time subspace integration for St. Venant-Kirchhoff deformable models}.
\newblock \bibinfo{journal}{\emph{ACM Trans. Graph.}} \bibinfo{volume}{24}, \bibinfo{number}{3} (\bibinfo{date}{July} \bibinfo{year}{2005}), \bibinfo{pages}{982–990}.
\newblock
\showISSN{0730-0301}
\urldef\tempurl%
\url{https://doi.org/10.1145/1073204.1073300}
\showDOI{\tempurl}


\bibitem[Benchekroun et~al\mbox{.}(2023)]%
        {benchekroun2023fastcomplementarydynamics}
\bibfield{author}{\bibinfo{person}{Otman Benchekroun}, \bibinfo{person}{Jiayi~Eris Zhang}, \bibinfo{person}{Siddartha Chaudhuri}, \bibinfo{person}{Eitan Grinspun}, \bibinfo{person}{Yi Zhou}, {and} \bibinfo{person}{Alec Jacobson}.} \bibinfo{year}{2023}\natexlab{}.
\newblock \showarticletitle{Fast Complementary Dynamics via Skinning Eigenmodes}.
\newblock \bibinfo{journal}{\emph{ACM Trans. Graph.}} \bibinfo{volume}{42}, \bibinfo{number}{4}, Article \bibinfo{articleno}{106} (\bibinfo{date}{July} \bibinfo{year}{2023}), \bibinfo{numpages}{21}~pages.
\newblock
\showISSN{0730-0301}
\urldef\tempurl%
\url{https://doi.org/10.1145/3592404}
\showDOI{\tempurl}


\bibitem[Bouaziz et~al\mbox{.}(2014)]%
        {bouaziz2014projectivedynamics}
\bibfield{author}{\bibinfo{person}{Sofien Bouaziz}, \bibinfo{person}{Sebastian Martin}, \bibinfo{person}{Tiantian Liu}, \bibinfo{person}{Ladislav Kavan}, {and} \bibinfo{person}{Mark Pauly}.} \bibinfo{year}{2014}\natexlab{}.
\newblock \showarticletitle{Projective dynamics: fusing constraint projections for fast simulation}.
\newblock \bibinfo{journal}{\emph{ACM Trans. Graph.}} \bibinfo{volume}{33}, \bibinfo{number}{4}, Article \bibinfo{articleno}{154} (\bibinfo{date}{July} \bibinfo{year}{2014}), \bibinfo{numpages}{11}~pages.
\newblock
\showISSN{0730-0301}
\urldef\tempurl%
\url{https://doi.org/10.1145/2601097.2601116}
\showDOI{\tempurl}


\bibitem[Brandt et~al\mbox{.}(2018)]%
        {brandt2018hyperreducedprojectivedynamics}
\bibfield{author}{\bibinfo{person}{Christopher Brandt}, \bibinfo{person}{Elmar Eisemann}, {and} \bibinfo{person}{Klaus Hildebrandt}.} \bibinfo{year}{2018}\natexlab{}.
\newblock \showarticletitle{Hyper-reduced projective dynamics}.
\newblock \bibinfo{journal}{\emph{ACM Trans. Graph.}} \bibinfo{volume}{37}, \bibinfo{number}{4}, Article \bibinfo{articleno}{80} (\bibinfo{date}{July} \bibinfo{year}{2018}), \bibinfo{numpages}{13}~pages.
\newblock
\showISSN{0730-0301}
\urldef\tempurl%
\url{https://doi.org/10.1145/3197517.3201387}
\showDOI{\tempurl}


\bibitem[Chang et~al\mbox{.}(2023)]%
        {changyue2023licrom}
\bibfield{author}{\bibinfo{person}{Yue Chang}, \bibinfo{person}{Peter~Yichen Chen}, \bibinfo{person}{Zhecheng Wang}, \bibinfo{person}{Maurizio~M. Chiaramonte}, \bibinfo{person}{Kevin Carlberg}, {and} \bibinfo{person}{Eitan Grinspun}.} \bibinfo{year}{2023}\natexlab{}.
\newblock \showarticletitle{LiCROM: Linear-Subspace Continuous Reduced Order Modeling with Neural Fields}. In \bibinfo{booktitle}{\emph{SIGGRAPH Asia 2023 Conference Papers}} (Sydney, NSW, Australia) \emph{(\bibinfo{series}{SA '23})}. \bibinfo{publisher}{Association for Computing Machinery}, \bibinfo{address}{New York, NY, USA}, Article \bibinfo{articleno}{111}, \bibinfo{numpages}{12}~pages.
\newblock
\showISBNx{9798400703157}
\urldef\tempurl%
\url{https://doi.org/10.1145/3610548.3618158}
\showDOI{\tempurl}


\bibitem[Chen et~al\mbox{.}(2024)]%
        {chen2024fluid}
\bibfield{author}{\bibinfo{person}{Yixin Chen}, \bibinfo{person}{David Levin}, {and} \bibinfo{person}{Timothy Langlois}.} \bibinfo{year}{2024}\natexlab{}.
\newblock \showarticletitle{Fluid Control with Laplacian Eigenfunctions}. In \bibinfo{booktitle}{\emph{ACM SIGGRAPH 2024 Conference Papers}}. \bibinfo{pages}{1--11}.
\newblock


\bibitem[Choi and Ko(2005)]%
        {choi2005modalwarping}
\bibfield{author}{\bibinfo{person}{Min~Gyu Choi} {and} \bibinfo{person}{Hyeong-Seok Ko}.} \bibinfo{year}{2005}\natexlab{}.
\newblock \showarticletitle{Modal Warping: Real-Time Simulation of Large Rotational Deformation and Manipulation}.
\newblock \bibinfo{journal}{\emph{IEEE Transactions on Visualization and Computer Graphics}} \bibinfo{volume}{11}, \bibinfo{number}{1} (\bibinfo{date}{Jan.} \bibinfo{year}{2005}), \bibinfo{pages}{91–101}.
\newblock
\showISSN{1077-2626}
\urldef\tempurl%
\url{https://doi.org/10.1109/TVCG.2005.13}
\showDOI{\tempurl}


\bibitem[Cui et~al\mbox{.}(2018)]%
        {cui2018laplacianeigenfluids}
\bibfield{author}{\bibinfo{person}{Qiaodong Cui}, \bibinfo{person}{Pradeep Sen}, {and} \bibinfo{person}{Theodore Kim}.} \bibinfo{year}{2018}\natexlab{}.
\newblock \showarticletitle{Scalable laplacian eigenfluids}.
\newblock \bibinfo{journal}{\emph{ACM Trans. Graph.}} \bibinfo{volume}{37}, \bibinfo{number}{4}, Article \bibinfo{articleno}{87} (\bibinfo{date}{July} \bibinfo{year}{2018}), \bibinfo{numpages}{12}~pages.
\newblock
\showISSN{0730-0301}
\urldef\tempurl%
\url{https://doi.org/10.1145/3197517.3201352}
\showDOI{\tempurl}


\bibitem[El-Atab et~al\mbox{.}(2020)]%
        {elatab2020soft}
\bibfield{author}{\bibinfo{person}{Nazek El-Atab}, \bibinfo{person}{Rishabh~B Mishra}, \bibinfo{person}{Fhad Al-Modaf}, \bibinfo{person}{Lana Joharji}, \bibinfo{person}{Aljohara~A Alsharif}, \bibinfo{person}{Haneen Alamoudi}, \bibinfo{person}{Marlon Diaz}, \bibinfo{person}{Nadeem Qaiser}, {and} \bibinfo{person}{Muhammad~Mustafa Hussain}.} \bibinfo{year}{2020}\natexlab{}.
\newblock \showarticletitle{Soft actuators for soft robotic applications: A review}.
\newblock \bibinfo{journal}{\emph{Advanced Intelligent Systems}} \bibinfo{volume}{2}, \bibinfo{number}{10} (\bibinfo{year}{2020}), \bibinfo{pages}{2000128}.
\newblock


\bibitem[Faure et~al\mbox{.}(2011)]%
        {faure2011Sparsemeshlessmethods}
\bibfield{author}{\bibinfo{person}{Fran\c{c}ois Faure}, \bibinfo{person}{Benjamin Gilles}, \bibinfo{person}{Guillaume Bousquet}, {and} \bibinfo{person}{Dinesh~K. Pai}.} \bibinfo{year}{2011}\natexlab{}.
\newblock \showarticletitle{Sparse meshless models of complex deformable solids}.
\newblock \bibinfo{journal}{\emph{ACM Trans. Graph.}} \bibinfo{volume}{30}, \bibinfo{number}{4}, Article \bibinfo{articleno}{73} (\bibinfo{date}{July} \bibinfo{year}{2011}), \bibinfo{numpages}{10}~pages.
\newblock
\showISSN{0730-0301}
\urldef\tempurl%
\url{https://doi.org/10.1145/2010324.1964968}
\showDOI{\tempurl}


\bibitem[Geijtenbeek et~al\mbox{.}(2013)]%
        {geijtenbeek2013flexiblemuscle}
\bibfield{author}{\bibinfo{person}{Thomas Geijtenbeek}, \bibinfo{person}{Michiel van~de Panne}, {and} \bibinfo{person}{A.~Frank van~der Stappen}.} \bibinfo{year}{2013}\natexlab{}.
\newblock \showarticletitle{Flexible Muscle-Based Locomotion for Bipedal Creatures}.
\newblock \bibinfo{journal}{\emph{ACM Transactions on Graphics}} \bibinfo{volume}{32}, \bibinfo{number}{6} (\bibinfo{year}{2013}).
\newblock


\bibitem[Ghanem and Spanos(2003)]%
        {ghanem2003stochastic}
\bibfield{author}{\bibinfo{person}{Roger~G Ghanem} {and} \bibinfo{person}{Pol~D Spanos}.} \bibinfo{year}{2003}\natexlab{}.
\newblock \bibinfo{booktitle}{\emph{Stochastic finite elements: a spectral approach}}.
\newblock \bibinfo{publisher}{Courier Corporation}.
\newblock


\bibitem[Hahn et~al\mbox{.}(2012)]%
        {hahn2012rigspacephysics}
\bibfield{author}{\bibinfo{person}{Fabian Hahn}, \bibinfo{person}{Sebastian Martin}, \bibinfo{person}{Bernhard Thomaszewski}, \bibinfo{person}{Robert Sumner}, \bibinfo{person}{Stelian Coros}, {and} \bibinfo{person}{Markus Gross}.} \bibinfo{year}{2012}\natexlab{}.
\newblock \showarticletitle{Rig-space physics}.
\newblock \bibinfo{journal}{\emph{ACM Trans. Graph.}} \bibinfo{volume}{31}, \bibinfo{number}{4}, Article \bibinfo{articleno}{72} (\bibinfo{date}{July} \bibinfo{year}{2012}), \bibinfo{numpages}{8}~pages.
\newblock
\showISSN{0730-0301}
\urldef\tempurl%
\url{https://doi.org/10.1145/2185520.2185568}
\showDOI{\tempurl}


\bibitem[Harmon and Zorin(2013)]%
        {harmon2013subspaceintegration}
\bibfield{author}{\bibinfo{person}{David Harmon} {and} \bibinfo{person}{Denis Zorin}.} \bibinfo{year}{2013}\natexlab{}.
\newblock \showarticletitle{Subspace integration with local deformations}.
\newblock \bibinfo{journal}{\emph{ACM Trans. Graph.}} \bibinfo{volume}{32}, \bibinfo{number}{4}, Article \bibinfo{articleno}{107} (\bibinfo{date}{July} \bibinfo{year}{2013}), \bibinfo{numpages}{10}~pages.
\newblock
\showISSN{0730-0301}
\urldef\tempurl%
\url{https://doi.org/10.1145/2461912.2461922}
\showDOI{\tempurl}


\bibitem[Hotelling(1933)]%
        {hotelling1933analysis}
\bibfield{author}{\bibinfo{person}{Harold Hotelling}.} \bibinfo{year}{1933}\natexlab{}.
\newblock \showarticletitle{Analysis of a complex of statistical variables into principal components.}
\newblock \bibinfo{journal}{\emph{Journal of educational psychology}} \bibinfo{volume}{24}, \bibinfo{number}{6} (\bibinfo{year}{1933}), \bibinfo{pages}{417}.
\newblock


\bibitem[Huang et~al\mbox{.}(2011)]%
        {huang2011RScoordantes}
\bibfield{author}{\bibinfo{person}{Jin Huang}, \bibinfo{person}{Yiying Tong}, \bibinfo{person}{Kun Zhou}, \bibinfo{person}{Hujun Bao}, {and} \bibinfo{person}{Mathieu Desbrun}.} \bibinfo{year}{2011}\natexlab{}.
\newblock \showarticletitle{Interactive Shape Interpolation through Controllable Dynamic Deformation}.
\newblock \bibinfo{journal}{\emph{IEEE Transactions on Visualization and Computer Graphics}} \bibinfo{volume}{17}, \bibinfo{number}{7} (\bibinfo{date}{July} \bibinfo{year}{2011}), \bibinfo{pages}{983–992}.
\newblock
\showISSN{1077-2626}
\urldef\tempurl%
\url{https://doi.org/10.1109/TVCG.2010.109}
\showDOI{\tempurl}


\bibitem[Jacobson et~al\mbox{.}(2012)]%
        {jacobson2012fastautomaticskinningtransformations}
\bibfield{author}{\bibinfo{person}{Alec Jacobson}, \bibinfo{person}{Ilya Baran}, \bibinfo{person}{Ladislav Kavan}, \bibinfo{person}{Jovan Popovi\'{c}}, {and} \bibinfo{person}{Olga Sorkine}.} \bibinfo{year}{2012}\natexlab{}.
\newblock \showarticletitle{Fast automatic skinning transformations}.
\newblock \bibinfo{journal}{\emph{ACM Trans. Graph.}} \bibinfo{volume}{31}, \bibinfo{number}{4}, Article \bibinfo{articleno}{77} (\bibinfo{date}{July} \bibinfo{year}{2012}), \bibinfo{numpages}{10}~pages.
\newblock
\showISSN{0730-0301}
\urldef\tempurl%
\url{https://doi.org/10.1145/2185520.2185573}
\showDOI{\tempurl}


\bibitem[Jacobson et~al\mbox{.}(2011)]%
        {jacobson2011boundedbiharmonicweights}
\bibfield{author}{\bibinfo{person}{Alec Jacobson}, \bibinfo{person}{Ilya Baran}, \bibinfo{person}{Jovan Popovi\'{c}}, {and} \bibinfo{person}{Olga Sorkine}.} \bibinfo{year}{2011}\natexlab{}.
\newblock \showarticletitle{Bounded biharmonic weights for real-time deformation}.
\newblock \bibinfo{journal}{\emph{ACM Trans. Graph.}} \bibinfo{volume}{30}, \bibinfo{number}{4}, Article \bibinfo{articleno}{78} (\bibinfo{date}{July} \bibinfo{year}{2011}), \bibinfo{numpages}{8}~pages.
\newblock
\showISSN{0730-0301}
\urldef\tempurl%
\url{https://doi.org/10.1145/2010324.1964973}
\showDOI{\tempurl}


\bibitem[James and Pai(2002)]%
        {james2002DyRt}
\bibfield{author}{\bibinfo{person}{Doug~L. James} {and} \bibinfo{person}{Dinesh~K. Pai}.} \bibinfo{year}{2002}\natexlab{}.
\newblock \showarticletitle{DyRT: dynamic response textures for real time deformation simulation with graphics hardware}.
\newblock \bibinfo{journal}{\emph{ACM Trans. Graph.}} \bibinfo{volume}{21}, \bibinfo{number}{3} (\bibinfo{date}{July} \bibinfo{year}{2002}), \bibinfo{pages}{582–585}.
\newblock
\showISSN{0730-0301}
\urldef\tempurl%
\url{https://doi.org/10.1145/566654.566621}
\showDOI{\tempurl}


\bibitem[Joshi et~al\mbox{.}(2007)]%
        {pushkar2007harmoniccoordinatesforcharacterarticulation}
\bibfield{author}{\bibinfo{person}{Pushkar Joshi}, \bibinfo{person}{Mark Meyer}, \bibinfo{person}{Tony DeRose}, \bibinfo{person}{Brian Green}, {and} \bibinfo{person}{Tom Sanocki}.} \bibinfo{year}{2007}\natexlab{}.
\newblock \showarticletitle{Harmonic coordinates for character articulation}.
\newblock \bibinfo{journal}{\emph{ACM Trans. Graph.}} \bibinfo{volume}{26}, \bibinfo{number}{3} (\bibinfo{date}{July} \bibinfo{year}{2007}), \bibinfo{pages}{71–es}.
\newblock
\showISSN{0730-0301}
\urldef\tempurl%
\url{https://doi.org/10.1145/1276377.1276466}
\showDOI{\tempurl}


\bibitem[Kim et~al\mbox{.}(2013)]%
        {kim2013nonlineardualmodes}
\bibfield{author}{\bibinfo{person}{Kwangkeun Kim}, \bibinfo{person}{Adrian~G Radu}, \bibinfo{person}{XQ Wang}, {and} \bibinfo{person}{Marc~P Mignolet}.} \bibinfo{year}{2013}\natexlab{}.
\newblock \showarticletitle{Nonlinear reduced order modeling of isotropic and functionally graded plates}.
\newblock \bibinfo{journal}{\emph{International Journal of Non-Linear Mechanics}}  \bibinfo{volume}{49} (\bibinfo{year}{2013}), \bibinfo{pages}{100--110}.
\newblock


\bibitem[Kim and Eberle(2022)]%
        {kim2022dynamicdeformables}
\bibfield{author}{\bibinfo{person}{Theodore Kim} {and} \bibinfo{person}{David Eberle}.} \bibinfo{year}{2022}\natexlab{}.
\newblock \showarticletitle{Dynamic deformables: implementation and production practicalities (now with code!)}. In \bibinfo{booktitle}{\emph{ACM SIGGRAPH 2022 Courses}} (Vancouver, British Columbia, Canada) \emph{(\bibinfo{series}{SIGGRAPH '22})}. \bibinfo{publisher}{Association for Computing Machinery}, \bibinfo{address}{New York, NY, USA}, Article \bibinfo{articleno}{7}, \bibinfo{numpages}{259}~pages.
\newblock
\showISBNx{9781450393621}
\urldef\tempurl%
\url{https://doi.org/10.1145/3532720.3535628}
\showDOI{\tempurl}


\bibitem[Kim and James(2011)]%
        {kim2011multidomain}
\bibfield{author}{\bibinfo{person}{Theodore Kim} {and} \bibinfo{person}{Doug~L. James}.} \bibinfo{year}{2011}\natexlab{}.
\newblock \showarticletitle{Physics-based character skinning using multi-domain subspace deformations}. In \bibinfo{booktitle}{\emph{Proceedings of the 2011 ACM SIGGRAPH/Eurographics Symposium on Computer Animation}} (Vancouver, British Columbia, Canada) \emph{(\bibinfo{series}{SCA '11})}. \bibinfo{publisher}{Association for Computing Machinery}, \bibinfo{address}{New York, NY, USA}, \bibinfo{pages}{63–72}.
\newblock
\showISBNx{9781450309233}
\urldef\tempurl%
\url{https://doi.org/10.1145/2019406.2019415}
\showDOI{\tempurl}


\bibitem[Lan et~al\mbox{.}(2021)]%
        {lan2021medialipc}
\bibfield{author}{\bibinfo{person}{Lei Lan}, \bibinfo{person}{Yin Yang}, \bibinfo{person}{Danny Kaufman}, \bibinfo{person}{Junfeng Yao}, \bibinfo{person}{Minchen Li}, {and} \bibinfo{person}{Chenfanfu Jiang}.} \bibinfo{year}{2021}\natexlab{}.
\newblock \showarticletitle{Medial IPC: accelerated incremental potential contact with medial elastics}.
\newblock \bibinfo{journal}{\emph{ACM Trans. Graph.}} \bibinfo{volume}{40}, \bibinfo{number}{4}, Article \bibinfo{articleno}{158} (\bibinfo{date}{July} \bibinfo{year}{2021}), \bibinfo{numpages}{16}~pages.
\newblock
\showISSN{0730-0301}
\urldef\tempurl%
\url{https://doi.org/10.1145/3450626.3459753}
\showDOI{\tempurl}


\bibitem[Langlois et~al\mbox{.}(2016)]%
        {langlois2016stocasticfem}
\bibfield{author}{\bibinfo{person}{Timothy Langlois}, \bibinfo{person}{Ariel Shamir}, \bibinfo{person}{Daniel Dror}, \bibinfo{person}{Wojciech Matusik}, {and} \bibinfo{person}{David I.~W. Levin}.} \bibinfo{year}{2016}\natexlab{}.
\newblock \showarticletitle{Stochastic structural analysis for context-aware design and fabrication}.
\newblock \bibinfo{journal}{\emph{ACM Trans. Graph.}} \bibinfo{volume}{35}, \bibinfo{number}{6}, Article \bibinfo{articleno}{226} (\bibinfo{date}{Dec.} \bibinfo{year}{2016}), \bibinfo{numpages}{13}~pages.
\newblock
\showISSN{0730-0301}
\urldef\tempurl%
\url{https://doi.org/10.1145/2980179.2982436}
\showDOI{\tempurl}


\bibitem[Lee et~al\mbox{.}(2018)]%
        {lee2018dexterousmanipulationvolumetricmuscles}
\bibfield{author}{\bibinfo{person}{Seunghwan Lee}, \bibinfo{person}{Ri Yu}, \bibinfo{person}{Jungnam Park}, \bibinfo{person}{Mridul Aanjaneya}, \bibinfo{person}{Eftychios Sifakis}, {and} \bibinfo{person}{Jehee Lee}.} \bibinfo{year}{2018}\natexlab{}.
\newblock \showarticletitle{Dexterous manipulation and control with volumetric muscles}.
\newblock \bibinfo{journal}{\emph{ACM Trans. Graph.}} \bibinfo{volume}{37}, \bibinfo{number}{4}, Article \bibinfo{articleno}{57} (\bibinfo{date}{July} \bibinfo{year}{2018}), \bibinfo{numpages}{13}~pages.
\newblock
\showISSN{0730-0301}
\urldef\tempurl%
\url{https://doi.org/10.1145/3197517.3201330}
\showDOI{\tempurl}


\bibitem[Li et~al\mbox{.}(2020)]%
        {Li2020IPC}
\bibfield{author}{\bibinfo{person}{Minchen Li}, \bibinfo{person}{Zachary Ferguson}, \bibinfo{person}{Teseo Schneider}, \bibinfo{person}{Timothy Langlois}, \bibinfo{person}{Denis Zorin}, \bibinfo{person}{Daniele Panozzo}, \bibinfo{person}{Chenfanfu Jiang}, {and} \bibinfo{person}{Danny~M. Kaufman}.} \bibinfo{year}{2020}\natexlab{}.
\newblock \showarticletitle{Incremental Potential Contact: Intersection- and Inversion-free Large Deformation Dynamics}.
\newblock \bibinfo{journal}{\emph{ACM Trans. Graph. (SIGGRAPH)}} \bibinfo{volume}{39}, \bibinfo{number}{4}, Article \bibinfo{articleno}{49} (\bibinfo{year}{2020}).
\newblock


\bibitem[Li et~al\mbox{.}(2019)]%
        {li2019optimizationtimeintegrator}
\bibfield{author}{\bibinfo{person}{Minchen Li}, \bibinfo{person}{Ming Gao}, \bibinfo{person}{Timothy Langlois}, \bibinfo{person}{Chenfanfu Jiang}, {and} \bibinfo{person}{Danny~M. Kaufman}.} \bibinfo{year}{2019}\natexlab{}.
\newblock \showarticletitle{Decomposed optimization time integrator for large-step elastodynamics}.
\newblock \bibinfo{journal}{\emph{ACM Trans. Graph.}} \bibinfo{volume}{38}, \bibinfo{number}{4}, Article \bibinfo{articleno}{70} (\bibinfo{date}{July} \bibinfo{year}{2019}), \bibinfo{numpages}{10}~pages.
\newblock
\showISSN{0730-0301}
\urldef\tempurl%
\url{https://doi.org/10.1145/3306346.3322951}
\showDOI{\tempurl}


\bibitem[Liu et~al\mbox{.}(2013)]%
        {liu2013fastsimofmasssprings}
\bibfield{author}{\bibinfo{person}{Tiantian Liu}, \bibinfo{person}{Adam~W. Bargteil}, \bibinfo{person}{James~F. O'Brien}, {and} \bibinfo{person}{Ladislav Kavan}.} \bibinfo{year}{2013}\natexlab{}.
\newblock \showarticletitle{Fast simulation of mass-spring systems}.
\newblock \bibinfo{journal}{\emph{ACM Trans. Graph.}} \bibinfo{volume}{32}, \bibinfo{number}{6}, Article \bibinfo{articleno}{214} (\bibinfo{date}{Nov.} \bibinfo{year}{2013}), \bibinfo{numpages}{7}~pages.
\newblock
\showISSN{0730-0301}
\urldef\tempurl%
\url{https://doi.org/10.1145/2508363.2508406}
\showDOI{\tempurl}


\bibitem[Mercier-Aubin et~al\mbox{.}(2022)]%
        {mercier-aubin2022adaptiverigidification}
\bibfield{author}{\bibinfo{person}{Alexandre Mercier-Aubin}, \bibinfo{person}{Paul~G. Kry}, \bibinfo{person}{Alexandre Winter}, {and} \bibinfo{person}{David I.~W. Levin}.} \bibinfo{year}{2022}\natexlab{}.
\newblock \showarticletitle{Adaptive Rigidification of Elastic Solids}.
\newblock \bibinfo{journal}{\emph{ACM Trans. Graph.}} \bibinfo{volume}{41}, \bibinfo{number}{4}, Article \bibinfo{articleno}{71}, \bibinfo{numpages}{11}~pages.
\newblock
\showISSN{0730-0301}
\urldef\tempurl%
\url{https://doi.org/10.1145/3528223.3530124}
\showDOI{\tempurl}


\bibitem[Min et~al\mbox{.}(2019)]%
        {min2019softcon}
\bibfield{author}{\bibinfo{person}{Sehee Min}, \bibinfo{person}{Jungdam Won}, \bibinfo{person}{Seunghwan Lee}, \bibinfo{person}{Jungnam Park}, {and} \bibinfo{person}{Jehee Lee}.} \bibinfo{year}{2019}\natexlab{}.
\newblock \showarticletitle{SoftCon: simulation and control of soft-bodied animals with biomimetic actuators}.
\newblock \bibinfo{journal}{\emph{ACM Trans. Graph.}} \bibinfo{volume}{38}, \bibinfo{number}{6}, Article \bibinfo{articleno}{208} (\bibinfo{date}{Nov.} \bibinfo{year}{2019}), \bibinfo{numpages}{12}~pages.
\newblock
\showISSN{0730-0301}
\urldef\tempurl%
\url{https://doi.org/10.1145/3355089.3356497}
\showDOI{\tempurl}


\bibitem[Modi et~al\mbox{.}(2021)]%
        {modi2021emu}
\bibfield{author}{\bibinfo{person}{V. Modi}, \bibinfo{person}{L. Fulton}, \bibinfo{person}{A. Jacobson}, \bibinfo{person}{S. Sueda}, {and} \bibinfo{person}{D.I.W. Levin}.} \bibinfo{year}{2021}\natexlab{}.
\newblock \showarticletitle{EMU: Efficient Muscle Simulation in Deformation Space}.
\newblock \bibinfo{journal}{\emph{Computer Graphics Forum}} \bibinfo{volume}{40}, \bibinfo{number}{1} (\bibinfo{year}{2021}), \bibinfo{pages}{234--248}.
\newblock
\urldef\tempurl%
\url{https://doi.org/10.1111/cgf.14185}
\showDOI{\tempurl}
\showeprint{https://onlinelibrary.wiley.com/doi/pdf/10.1111/cgf.14185}


\bibitem[Modi et~al\mbox{.}(2024)]%
        {modi2024simplicits}
\bibfield{author}{\bibinfo{person}{Vismay Modi}, \bibinfo{person}{Nicholas Sharp}, \bibinfo{person}{Or Perel}, \bibinfo{person}{Shinjiro Sueda}, {and} \bibinfo{person}{David I.~W. Levin}.} \bibinfo{year}{2024}\natexlab{}.
\newblock \showarticletitle{Simplicits: Mesh-Free, Geometry-Agnostic Elastic Simulation}.
\newblock \bibinfo{journal}{\emph{ACM Trans. Graph.}} \bibinfo{volume}{43}, \bibinfo{number}{4}, Article \bibinfo{articleno}{117} (\bibinfo{date}{July} \bibinfo{year}{2024}), \bibinfo{numpages}{11}~pages.
\newblock
\showISSN{0730-0301}
\urldef\tempurl%
\url{https://doi.org/10.1145/3658184}
\showDOI{\tempurl}


\bibitem[Nocedal and Wright(1999)]%
        {nocedal1999numerical}
\bibfield{author}{\bibinfo{person}{Jorge Nocedal} {and} \bibinfo{person}{Stephen~J Wright}.} \bibinfo{year}{1999}\natexlab{}.
\newblock \bibinfo{booktitle}{\emph{Numerical optimization}}.
\newblock \bibinfo{publisher}{Springer}.
\newblock


\bibitem[Pan et~al\mbox{.}(2015)]%
        {pan2015subspaceRS}
\bibfield{author}{\bibinfo{person}{Zherong Pan}, \bibinfo{person}{Hujun Bao}, {and} \bibinfo{person}{Jin Huang}.} \bibinfo{year}{2015}\natexlab{}.
\newblock \showarticletitle{Subspace dynamic simulation using rotation-strain coordinates}.
\newblock \bibinfo{journal}{\emph{ACM Trans. Graph.}} \bibinfo{volume}{34}, \bibinfo{number}{6}, Article \bibinfo{articleno}{242} (\bibinfo{date}{Nov.} \bibinfo{year}{2015}), \bibinfo{numpages}{12}~pages.
\newblock
\showISSN{0730-0301}
\urldef\tempurl%
\url{https://doi.org/10.1145/2816795.2818090}
\showDOI{\tempurl}


\bibitem[Pentland and Williams(1989)]%
        {pentlandwilliams1989goodvibrations}
\bibfield{author}{\bibinfo{person}{A. Pentland} {and} \bibinfo{person}{J. Williams}.} \bibinfo{year}{1989}\natexlab{}.
\newblock \showarticletitle{Good vibrations: modal dynamics for graphics and animation}. In \bibinfo{booktitle}{\emph{Proceedings of the 16th Annual Conference on Computer Graphics and Interactive Techniques}} \emph{(\bibinfo{series}{SIGGRAPH '89})}. \bibinfo{publisher}{Association for Computing Machinery}, \bibinfo{address}{New York, NY, USA}, \bibinfo{pages}{215–222}.
\newblock
\showISBNx{0897913124}
\urldef\tempurl%
\url{https://doi.org/10.1145/74333.74355}
\showDOI{\tempurl}


\bibitem[Sharp et~al\mbox{.}(2023)]%
        {sharp2023datafreelearningreducedorderkinematics}
\bibfield{author}{\bibinfo{person}{Nicholas Sharp}, \bibinfo{person}{Cristian Romero}, \bibinfo{person}{Alec Jacobson}, \bibinfo{person}{Etienne Vouga}, \bibinfo{person}{Paul Kry}, \bibinfo{person}{David~I.W. Levin}, {and} \bibinfo{person}{Justin Solomon}.} \bibinfo{year}{2023}\natexlab{}.
\newblock \showarticletitle{Data-Free Learning of Reduced-Order Kinematics}. In \bibinfo{booktitle}{\emph{ACM SIGGRAPH 2023 Conference Proceedings}} (Los Angeles, CA, USA) \emph{(\bibinfo{series}{SIGGRAPH '23})}. \bibinfo{publisher}{Association for Computing Machinery}, \bibinfo{address}{New York, NY, USA}, Article \bibinfo{articleno}{40}, \bibinfo{numpages}{9}~pages.
\newblock
\showISBNx{9798400701597}
\urldef\tempurl%
\url{https://doi.org/10.1145/3588432.3591521}
\showDOI{\tempurl}


\bibitem[Sifakis and Barbič(2012)]%
        {sifakis2012notes}
\bibfield{author}{\bibinfo{person}{Eftychios Sifakis} {and} \bibinfo{person}{Jernej Barbič}.} \bibinfo{year}{2012}\natexlab{}.
\newblock \showarticletitle{FEM Simulation of 3D Deformable Solids: A Practitioner's Guide to Theory, Discretization and Model Reduction}. In \bibinfo{booktitle}{\emph{ACM SIGGRAPH 2012 Courses}} (Los Angeles, California, USA). \bibinfo{publisher}{Association for Computing Machinery}, \bibinfo{address}{New York, NY, USA}, \bibinfo{pages}{20:1--20:50}.
\newblock
\showISBNx{978-1-4503-1678-1}
\urldef\tempurl%
\url{https://doi.org/10.1145/2343483.2343501}
\showDOI{\tempurl}


\bibitem[Sirovich(1987)]%
        {sirovich1987turbulence}
\bibfield{author}{\bibinfo{person}{Lawrence Sirovich}.} \bibinfo{year}{1987}\natexlab{}.
\newblock \showarticletitle{Turbulence and the dynamics of coherent structures. I. Coherent structures}.
\newblock \bibinfo{journal}{\emph{Quarterly of applied mathematics}} \bibinfo{volume}{45}, \bibinfo{number}{3} (\bibinfo{year}{1987}), \bibinfo{pages}{561--571}.
\newblock


\bibitem[Tan et~al\mbox{.}(2012)]%
        {tan2012softbodylocomotion}
\bibfield{author}{\bibinfo{person}{Jie Tan}, \bibinfo{person}{Greg Turk}, {and} \bibinfo{person}{C.~Karen Liu}.} \bibinfo{year}{2012}\natexlab{}.
\newblock \showarticletitle{Soft body locomotion}.
\newblock \bibinfo{journal}{\emph{ACM Trans. Graph.}} \bibinfo{volume}{31}, \bibinfo{number}{4}, Article \bibinfo{articleno}{26} (\bibinfo{date}{July} \bibinfo{year}{2012}), \bibinfo{numpages}{11}~pages.
\newblock
\showISSN{0730-0301}
\urldef\tempurl%
\url{https://doi.org/10.1145/2185520.2185522}
\showDOI{\tempurl}


\bibitem[Teng et~al\mbox{.}(2015)]%
        {yun2015subspacecondensation}
\bibfield{author}{\bibinfo{person}{Yun Teng}, \bibinfo{person}{Mark Meyer}, \bibinfo{person}{Tony DeRose}, {and} \bibinfo{person}{Theodore Kim}.} \bibinfo{year}{2015}\natexlab{}.
\newblock \showarticletitle{Subspace condensation: full space adaptivity for subspace deformations}.
\newblock \bibinfo{journal}{\emph{ACM Trans. Graph.}} \bibinfo{volume}{34}, \bibinfo{number}{4}, Article \bibinfo{articleno}{76} (\bibinfo{date}{July} \bibinfo{year}{2015}), \bibinfo{numpages}{9}~pages.
\newblock
\showISSN{0730-0301}
\urldef\tempurl%
\url{https://doi.org/10.1145/2766904}
\showDOI{\tempurl}


\bibitem[Teran et~al\mbox{.}(2003)]%
        {teran2003skeletalmuscle}
\bibfield{author}{\bibinfo{person}{J. Teran}, \bibinfo{person}{S. Blemker}, \bibinfo{person}{V.~Ng~Thow Hing}, {and} \bibinfo{person}{R. Fedkiw}.} \bibinfo{year}{2003}\natexlab{}.
\newblock \showarticletitle{Finite volume methods for the simulation of skeletal muscle}. In \bibinfo{booktitle}{\emph{Proceedings of the 2003 ACM SIGGRAPH/Eurographics Symposium on Computer Animation}} (San Diego, California) \emph{(\bibinfo{series}{SCA '03})}. \bibinfo{publisher}{Eurographics Association}, \bibinfo{address}{Goslar, DEU}, \bibinfo{pages}{68–74}.
\newblock
\showISBNx{1581136595}


\bibitem[Trusty et~al\mbox{.}(2023)]%
        {trusty2023subspacemfem}
\bibfield{author}{\bibinfo{person}{Ty Trusty}, \bibinfo{person}{Otman Benchekroun}, \bibinfo{person}{Eitan Grinspun}, \bibinfo{person}{Danny~M. Kaufman}, {and} \bibinfo{person}{David~I.W. Levin}.} \bibinfo{year}{2023}\natexlab{}.
\newblock \showarticletitle{Subspace Mixed Finite Elements for Real-Time Heterogeneous Elastodynamics}. In \bibinfo{booktitle}{\emph{SIGGRAPH Asia 2023 Conference Papers}} (Sydney, NSW, Australia) \emph{(\bibinfo{series}{SA '23})}. \bibinfo{publisher}{Association for Computing Machinery}, \bibinfo{address}{New York, NY, USA}, Article \bibinfo{articleno}{112}, \bibinfo{numpages}{10}~pages.
\newblock
\showISBNx{9798400703157}
\urldef\tempurl%
\url{https://doi.org/10.1145/3610548.3618220}
\showDOI{\tempurl}


\bibitem[Trusty et~al\mbox{.}(2024)]%
        {trusty2024tradingspaces}
\bibfield{author}{\bibinfo{person}{Ty Trusty}, \bibinfo{person}{Yun~(Raymond) Fei}, \bibinfo{person}{David Levin}, {and} \bibinfo{person}{Danny Kaufman}.} \bibinfo{year}{2024}\natexlab{}.
\newblock \showarticletitle{Trading Spaces: Adaptive Subspace Time Integration for Contacting Elastodynamics}.
\newblock \bibinfo{journal}{\emph{ACM Trans. Graph.}} \bibinfo{volume}{43}, \bibinfo{number}{6}, Article \bibinfo{articleno}{227} (\bibinfo{date}{Nov.} \bibinfo{year}{2024}), \bibinfo{numpages}{16}~pages.
\newblock
\showISSN{0730-0301}
\urldef\tempurl%
\url{https://doi.org/10.1145/3687946}
\showDOI{\tempurl}


\bibitem[Virtanen et~al\mbox{.}(2020)]%
        {scipy}
\bibfield{author}{\bibinfo{person}{Pauli Virtanen}, \bibinfo{person}{Ralf Gommers}, \bibinfo{person}{Travis~E. Oliphant}, \bibinfo{person}{Matt Haberland}, \bibinfo{person}{Tyler Reddy}, \bibinfo{person}{David Cournapeau}, \bibinfo{person}{Evgeni Burovski}, \bibinfo{person}{Pearu Peterson}, \bibinfo{person}{Warren Weckesser}, \bibinfo{person}{Jonathan Bright}, \bibinfo{person}{St{\'e}fan~J. {van der Walt}}, \bibinfo{person}{Matthew Brett}, \bibinfo{person}{Joshua Wilson}, \bibinfo{person}{K.~Jarrod Millman}, \bibinfo{person}{Nikolay Mayorov}, \bibinfo{person}{Andrew R.~J. Nelson}, \bibinfo{person}{Eric Jones}, \bibinfo{person}{Robert Kern}, \bibinfo{person}{Eric Larson}, \bibinfo{person}{C~J Carey}, \bibinfo{person}{{\.I}lhan Polat}, \bibinfo{person}{Yu Feng}, \bibinfo{person}{Eric~W. Moore}, \bibinfo{person}{Jake {VanderPlas}}, \bibinfo{person}{Denis Laxalde}, \bibinfo{person}{Josef Perktold}, \bibinfo{person}{Robert Cimrman}, \bibinfo{person}{Ian Henriksen}, \bibinfo{person}{E.~A. Quintero},
  \bibinfo{person}{Charles~R. Harris}, \bibinfo{person}{Anne~M. Archibald}, \bibinfo{person}{Ant{\^o}nio~H. Ribeiro}, \bibinfo{person}{Fabian Pedregosa}, \bibinfo{person}{Paul {van Mulbregt}}, {and} \bibinfo{person}{{SciPy 1.0 Contributors}}.} \bibinfo{year}{2020}\natexlab{}.
\newblock \showarticletitle{{{SciPy} 1.0: Fundamental Algorithms for Scientific Computing in Python}}.
\newblock \bibinfo{journal}{\emph{Nature Methods}}  \bibinfo{volume}{17} (\bibinfo{year}{2020}), \bibinfo{pages}{261--272}.
\newblock
\urldef\tempurl%
\url{https://doi.org/10.1038/s41592-019-0686-2}
\showDOI{\tempurl}


\bibitem[von Tycowicz et~al\mbox{.}(2013)]%
        {vontycowicz2013efficientreduceddeformableobjects}
\bibfield{author}{\bibinfo{person}{Christoph von Tycowicz}, \bibinfo{person}{Christian Schulz}, \bibinfo{person}{Hans-Peter Seidel}, {and} \bibinfo{person}{Klaus Hildebrandt}.} \bibinfo{year}{2013}\natexlab{}.
\newblock \showarticletitle{An efficient construction of reduced deformable objects}.
\newblock \bibinfo{journal}{\emph{ACM Trans. Graph.}} \bibinfo{volume}{32}, \bibinfo{number}{6}, Article \bibinfo{articleno}{213} (\bibinfo{date}{Nov.} \bibinfo{year}{2013}), \bibinfo{numpages}{10}~pages.
\newblock
\showISSN{0730-0301}
\urldef\tempurl%
\url{https://doi.org/10.1145/2508363.2508392}
\showDOI{\tempurl}


\bibitem[Wang et~al\mbox{.}(2015)]%
        {yu2015linearsubspacedesign}
\bibfield{author}{\bibinfo{person}{Yu Wang}, \bibinfo{person}{Alec Jacobson}, \bibinfo{person}{Jernej Barbi\v{c}}, {and} \bibinfo{person}{Ladislav Kavan}.} \bibinfo{year}{2015}\natexlab{}.
\newblock \showarticletitle{Linear subspace design for real-time shape deformation}.
\newblock \bibinfo{journal}{\emph{ACM Trans. Graph.}} \bibinfo{volume}{34}, \bibinfo{number}{4}, Article \bibinfo{articleno}{57} (\bibinfo{date}{July} \bibinfo{year}{2015}), \bibinfo{numpages}{11}~pages.
\newblock
\showISSN{0730-0301}
\urldef\tempurl%
\url{https://doi.org/10.1145/2766952}
\showDOI{\tempurl}


\bibitem[Wu and Umetani(2023)]%
        {wu2023twowaycouplingskinningtransformationspbd}
\bibfield{author}{\bibinfo{person}{Yuhan Wu} {and} \bibinfo{person}{Nobuyuki Umetani}.} \bibinfo{year}{2023}\natexlab{}.
\newblock \showarticletitle{Two-Way Coupling of Skinning Transformations and Position Based Dynamics}.
\newblock \bibinfo{journal}{\emph{Proc. ACM Comput. Graph. Interact. Tech.}} \bibinfo{volume}{6}, \bibinfo{number}{3}, Article \bibinfo{articleno}{47} (\bibinfo{date}{Aug.} \bibinfo{year}{2023}), \bibinfo{numpages}{18}~pages.
\newblock
\urldef\tempurl%
\url{https://doi.org/10.1145/3606930}
\showDOI{\tempurl}


\bibitem[Xavier et~al\mbox{.}(2022)]%
        {xavier2022soft}
\bibfield{author}{\bibinfo{person}{Matheus~S Xavier}, \bibinfo{person}{Charbel~D Tawk}, \bibinfo{person}{Ali Zolfagharian}, \bibinfo{person}{Joshua Pinskier}, \bibinfo{person}{David Howard}, \bibinfo{person}{Taylor Young}, \bibinfo{person}{Jiewen Lai}, \bibinfo{person}{Simon~M Harrison}, \bibinfo{person}{Yuen~K Yong}, \bibinfo{person}{Mahdi Bodaghi}, {et~al\mbox{.}}} \bibinfo{year}{2022}\natexlab{}.
\newblock \showarticletitle{Soft pneumatic actuators: A review of design, fabrication, modeling, sensing, control and applications}.
\newblock \bibinfo{journal}{\emph{IEEE Access}}  \bibinfo{volume}{10} (\bibinfo{year}{2022}), \bibinfo{pages}{59442--59485}.
\newblock


\bibitem[Xu and Barbi\v{c}(2016)]%
        {xu2016posespacesubspacedynamics}
\bibfield{author}{\bibinfo{person}{Hongyi Xu} {and} \bibinfo{person}{Jernej Barbi\v{c}}.} \bibinfo{year}{2016}\natexlab{}.
\newblock \showarticletitle{Pose-space subspace dynamics}.
\newblock \bibinfo{journal}{\emph{ACM Trans. Graph.}} \bibinfo{volume}{35}, \bibinfo{number}{4}, Article \bibinfo{articleno}{35} (\bibinfo{date}{July} \bibinfo{year}{2016}), \bibinfo{numpages}{14}~pages.
\newblock
\showISSN{0730-0301}
\urldef\tempurl%
\url{https://doi.org/10.1145/2897824.2925916}
\showDOI{\tempurl}


\bibitem[Zhang et~al\mbox{.}(2020)]%
        {zhang2020complementarydynamics}
\bibfield{author}{\bibinfo{person}{Jiayi~Eris Zhang}, \bibinfo{person}{Seungbae Bang}, \bibinfo{person}{David I.~W. Levin}, {and} \bibinfo{person}{Alec Jacobson}.} \bibinfo{year}{2020}\natexlab{}.
\newblock \showarticletitle{Complementary dynamics}.
\newblock \bibinfo{journal}{\emph{ACM Trans. Graph.}} \bibinfo{volume}{39}, \bibinfo{number}{6}, Article \bibinfo{articleno}{179} (\bibinfo{date}{Nov.} \bibinfo{year}{2020}), \bibinfo{numpages}{11}~pages.
\newblock
\showISSN{0730-0301}
\urldef\tempurl%
\url{https://doi.org/10.1145/3414685.3417819}
\showDOI{\tempurl}


\end{thebibliography}

\appendix 
\section{Deriving Modal Analysis from Force-Dual Modes}
\label{app:modal_analysis_from_force_dual_modes}

The force distribution has $\Sigma_F = \mat{M}$. Hence, we arrive at $\Sigma_U = \mat{H}^{-1} \mat{M} \mat{H}^{-1}$ from \cref{eq:statics_var_u}. The corresponding eigenvalue equation for the optimal basis $\mat{B}$ is the lowest eigenvectors of
\begin{equation}
    \mat{H} \mat{M}^{-1} \mat{H} \mat{B} = \mat{M} \mat{B} \Lambda^{-1} \,.
\end{equation}
Observe the modes $\mat{B}_{\text{modal}}$ satisfy this equation, since,
\begin{equation}
\begin{split}
    \mat{H} \mat{M}^{-1} (\mat{H} \mat{B}_{\text{modal}}) &= \mat{H} \mat{M}^{-1} \mat{M} \mat{B}_{\text{modal}} \mat{\Gamma} \\
    &= \mat{H} \mat{B}_{\text{modal}} \mat{\Gamma} =  \M \B_{\text{modal}} \mat{\Gamma}^2 \,,
\end{split}
\end{equation}
with $\mat{\Gamma}^2 = \Lambda^{-1}$. Hence, modal analysis arises naturally as the optimal linear basis for this force distribution.
\begin{equation}
 \alpha \H \mat{B} = \mat{M} \mat{B} \Lambda^{-1} \,.
\end{equation}
which we observe is the same as the modal analysis equations \cref{eq:modal_analysis} but with $\mat{\Gamma} = \alpha^{-1} \Lambda^{-1}$.

\section{Deriving Optimal Subspace reconstruction}
\label{app:deriving-optimal-subspace}

We can derive a mass orthonormal PCA/KKL as the subspace matrix $\B \in \mathbb{R}^{n \times m}$ that minimizes the reconstruction error for the stochastic process $\qoi$. 
\begin{align} \label{eq:optimization_best_space}
    \argmin_{\B} \mathbb{E} \left[ || \qoi - \hat \qoi||^2_\M  \right ] \quad s.t. \quad \B^T \M \B = \I 
\end{align}
Where $\hat \qoi = \B^\dag \qoi = \B \B^T \M \qoi$ is the projected approximation of the stochastic process. Note, we assume without loss of generality that $\qoi$ is mean-zero. First, consider the term inside the expectation,
\begin{equation}
\begin{split}
         \|\qoi - \hat \qoi\|_\M^2 &= \tr \left( \qoi^T \M \qoi + \hat \qoi^T \M  \hat  \qoi - 2 \hat \qoi^T \M  \qoi \right) \,.
\end{split}
\end{equation}
We drop the constant term $\qoi^T \M \qoi$ and continue,
\begin{equation}
\begin{split}
   \|\qoi - \hat \qoi\|_\M^2 &\cong \tr \left( \qoi^T \M \B \underbrace{\B^T \M  \B}_{\I} \B^T \M \qoi - 2 \qoi^T \M \B \B^T \M  \qoi \right) \\
    &= \tr \left( \qoi^T \M \B  \B^T \M \qoi - 2 \qoi^T \M \B \B^T \M  \qoi \right)  \\
    &= \tr \left( \B^T \M \qoi \qoi^T \M \B \right) - 2 \tr \left( \B^T  \M \qoi \qoi^T \M \B  \right) \\
    &= - \tr \left(   \B^T \M \qoi \qoi^T \M \B \right)\,.
\end{split}
\end{equation}
Taking expectations and interchanging with trace gives
\begin{equation}
    \E  \|\qoi - \hat \qoi\|_\M^2 \cong - \tr \left(   \B^T \M \E[\qoi \qoi^T] \M \B \right) = - \tr \left(   \B^T \M \Sigma_\qoi \M \B \right) \,.
\end{equation}

Introducing Lagrange multipliers $\Lambda$ for the constraint, the Lagrangian of \cref{eq:optimization_best_space} is written
\begin{equation}
    \mathcal{L} = -\tr \left(\B^T \M \Sigma_\qoi \M \B \right) + \tr(\Lambda  (\B^T \M \B  - \I)) \,.
\end{equation}
The variation of the Lagrangian is given by
\begin{equation}
\begin{split}
    \delta \mathcal{L} &= -2 \tr \left((\delta\B)^T \M \Sigma_\qoi \M \B \right) + 2 \tr(\Lambda ((\delta \B)^T \M \B)) \\
    &= -2 \tr \left((\delta\B)^T (\M \Sigma_\qoi \M \B - \M \B \Lambda)\right) \,,
\end{split}
\end{equation}
where we have used the symmetry of trace while taking variations. Now, the only way $\delta \mathcal{L}$ can be zero for all variations $\delta \B$ is for $\M \Sigma_\qoi \M \B - \M \B \Lambda$ to identically be zero. Thus, the critical points of the optimization satisfy the truncated generalized eigenvalue problem
\begin{align}
     \M \Sigma_\qoi \M \B = \M \B \Lambda \,,
\end{align}
with $\Lambda$ the largest eigenvalues of $\M \Sigma_\qoi \M$, and $\B$ their corresponding eigenvectors. 
This can be further simplified by multiplying both sides by $\M^{-1}$,
\begin{align} \label{eq:optimization_critical_point}
     \Sigma_\qoi \M \B = \B \Lambda \,.
\end{align}

While this tells us that $\B$ should satisfy the above eigenvalue equation, it doesn't tell us \emph{which} eigenvectors we should take. In order to answer that question, we plug \cref{eq:optimization_critical_point} back into the objective to see that
\begin{equation}
    \E  \|\qoi - \hat \qoi\|_\M^2 \cong- \tr \left(   \B^T \M \B \Lambda \right) = - \tr \Lambda \,.
\end{equation}
Hence, in order to minimize the objective, we want to choose the \emph{top} $m$ eigenvectors of the generalized eigenvalue problem \cref{eq:optimization_critical_point}.

\section{Deriving Green's Functions from Force-Dual Modes}
\label{app:greens_functions_from_force_dual_modes}

We consider the special case where the force prior is low-rank Gaussian, $F \sim \mathcal{N}(0, \mat{F}\mat{F}^\top)$. 
where $\F \in \mathbb{R}^{3n \times r}$ is a tall skinny of matrix of $r$ different applied forces.
 This force distribution induces a displacement distribution with covariance $\mat{\Sigma}_{U} = \mat{G} \mat{G}^\top$, where $\mat{G} = \mat{H}^{-1}\mat{F}$. %
The Force-Dual framework constructs a subspace by solving the generalized eigenvalue problem
$\mat{\Sigma}_{U} \mat{M} \mat{B} = \mat{B} \mat{\Lambda},$
which becomes
\[
\mat{G} \mat{G}^\top \mat{M} \mat{B} = \mat{B} \mat{\Lambda} \ .
\]

Examining the left hand side: Since for any matrix $\mat{A}$, $\textrm{Col}(\mat{G}\mat{A}) \subset \textrm{Col}(\mat{G})$,
therefore $\textrm{Col}(\textrm{LHS}) \subset \textrm{Col}(\mat{G})$. 
Turning to the RHS, since $\Lambda$ is invertible, $\textrm{Col}(\mat{B}) = \textrm{Col}(\mat{B}\mat{\Lambda}) = \textrm{Col}(\textrm{RHS})$. 
Finally, since $\textrm{Col}(\textrm{LHS})=\textrm{Col}(\textrm{RHS})$, therefore
$\textrm{Col}(\mat{B}) = \textrm{Col}(\textrm{LHS}) \subset \textrm{Col}(\mat{G})$.

\end{document}